\documentclass[fleqn,usenatbib]{mnras}

\usepackage{newtxtext,newtxmath}
\usepackage[T1]{fontenc}
\usepackage{orcidlink}
\usepackage{academicons}
\usepackage{changes}

\newcommand{\orcid}[1]{\href{https://orcid.org/#1}{\textcolor[HTML]{A6CE39}{\aiOrcid}}}

\newcommand{\nb}{\textsc{Nbody6++GPU}}

\DeclareRobustCommand{\VAN}[3]{#2}
\let\VANthebibliography\thebibliography
\def\thebibliography{\DeclareRobustCommand{\VAN}[3]{##3}\VANthebibliography}

\usepackage{graphicx}	% Including figure files
\usepackage{amsmath}	% Advanced maths commands
\usepackage{amssymb}	% Extra maths symbols
\usepackage{xcolor} 

\title[Hot Jupiter Formation in dense Star Clusters]{Hot Jupiter Formation in Dense Star Clusters}

\author[L. Benkendorff et al.]{
L. Benkendorff$^{1}$\orcidlink{0009-0008-8088-2918}\thanks{Leonard.Benkendorff@gmail.com},
F. Flammini Dotti$^{1}$\orcidlink{0000-0002-8881-3078},
K. Stock$^{1,2}$\orcidlink{0000-0003-0321-3709},
Maxwell.X. Cai $^{5}$\orcidlink{0000-0002-1116-2705}, 
and R. Spurzem $^{1,3,4}$\orcidlink{0000-0003-2264-7203}
\\
$^{1}$Astronomisches Rechen-Institut, Zentrum f\"ur Astronomie der Universität Heidelberg, Mönchhofstraße 12-14, D-69120 Heidelberg,
Germany\\
$^{2}$ GMV, Europaplatz 2, 64293 Darmstadt, Germany \\
$^{3}$ National Astronomical Observatories and Key Laboratory of Computational Astrophysics, Chinese Academy of Sciences, \\20A Datun Rd.,Chaoyang District, 100101, Beijing, China\\
$^{4}$ Kavli Institute for Astronomy and Astrophysics, Peking University, 5 Yi He Yuan Road, Haidian District, Beijing 100871, P.R. China \\
$^{5}$ Leiden Observatory, Leiden University, P.O. Box 9513, 2300 RA, Leiden, The Netherlands 
}

\date{Accepted 2024 January 13. Received 2023 December 30; in original form 2023 August 21}

\pubyear{2023}

\begin{document}
\label{firstpage}
\pagerange{\pageref{firstpage}--\pageref{lastpage}}
\maketitle

\begin{abstract}
Hot Jupiters (HJ) are defined as Jupiter-mass exoplanets orbiting around their host star with an orbital period < 10\,days. It is assumed that HJ do not form in-situ but ex-situ. Recent discoveries show that star clusters contribute to the formation of HJ. 
We present direct $N$-body simulations of planetary systems in star clusters and analyze the formation of HJ in them. We combine two direct $N$-body codes: \nb{} for the dynamics of dense star clusters with 32 000 and 64 000 stellar members and \textsc{LonelyPlanets} used to follow 200 identical planetary systems around solar mass stars in those star clusters. We use different sets with 3, 4, or 5 planets and with the innermost planet at a semi-major axis of 5\,au or 1\,au and follow them for 100 Myr in our simulations. The results indicate that HJs are generated with high efficiency in dense star clusters if the innermost planet is already close to the host star at a semi-major axis of 1\,au.
If the innermost planet is initially beyond a semi-major axis of 5\,au, the probability of a potential HJ ranges between $1.5-4.5$ percent. Very dense stellar neighborhoods tend to eject planets rather than forming HJs. A correlation between HJ formation and angular momentum deficit (AMD) is not witnessed. Young Hot Jupiters ($t_{\rm age} < 100$ Myrs) have only been found, in our simulations, in planetary systems with the innermost planet at a semi-major axis of 1\,au. 
\end{abstract}


%%%%%%%%%%%%%%%%%%%%%%%%%%%%%%%%%%%%%%%%%%%%%%%%%%

%%%%%%%%%%%%%%%%% BODY OF PAPER %%%%%%%%%%%%%%%%%%

\section{Introduction}
Hot Jupiters (hereafter HJ) are exoplanets that have a mass of $\geq 0.25$ Jupiter mass and orbit a star with an orbital period of less than 10 days \citep[e.g.,][]{Dawson2018}. So far, $\approx 600$ HJs have been confirmed\footnote{\url{exoplanetarchive.ipac.caltech.edu} accessed on Dec. 2023.}. Because of their relatively short orbits and large masses, they are ideal targets for detection by the radial velocity method as well as the transit method \citep[e.g.,][]{Dawson2018}. Hot Jupiters have been found around Solar-type stars at a rate of about 1\,\% \citep[e.g.,][]{Dawson2018}. Their origin is not yet fully understood - three main theories are discussed: in-situ formation, primordial disc migration, and high-eccentricity migration.\\
The \textit{in-situ} model claims that Hot Jupiters are formed via core accretion close to their current location \citep{Bodenheimer2000, Batygin2016, Boley2016}. However, it has been argued that the primordial mass budget of protoplanetary material is not sufficient in the hot regions close to the star to form a sufficiently large core \citep[e.g.,][]{Lin1996}. \\
Therefore, the primordial disc migration model suggests that Hot Jupiters form beyond the snow-line, where icy materials are abundant \citep[e.g.,][]{Walsh2011,Dawson2018} and migrate inwards thereafter. The first model regarding the formation of Hot Jupiters suggests that planets are losing angular momentum due to the interaction with the protoplanetary disk \citep{Lin1996, Masset2003, Ida2008}. The planet has to reach its final destination within the lifetime of the disk. However, the model struggles to explain Hot Jupiters with eccentric orbits \citep[e.g.,][]{Rouan2012} or a misalignment between the orbital angular momentum of the planet and the spin of the host star \citep[e.g.,][]{Triaud2010}. Furthermore, \citet{Winter2020} show that Hot Jupiter occurrence is correlated with stellar clustering, which is not consistent with that model. \\
Most stars are born in star clusters with up to a few thousand members \citep[e.g.,][]{Lada2003}. It is plausible that stellar encounters in the early stages significantly influence the dynamics of planetary systems. Numerical simulations have shown that planets around members of dense star clusters are likely to be ejected from their planetary system \citep[]{Spurzem2009, Cai2017, Cai2018, Dotti2019, Stock2020, Wu2023}. Only less than 1\% of known exoplanets are found in star clusters \citep{Cai2019}.  On the other hand, Hot Jupiter formation in a star cluster environment might be favorable, as the high rate of Hot Jupiters in the old open star cluster M67 \citep{Brucalassi2017} and the mentioned correlation of stellar density and Hot Jupiter occurrence \citep[e.g.,][]{Winter2020} suggest. \\
Gravitational interactions with other planets or stars can excite
high eccentricity for the orbit of a Jupiter-mass object \citep[e.g.,][]{Spurzem2009, Dotti2019}. Then tidal interaction with the host star would decrease the semi-major axis and eccentricity and eventually form a Hot Jupiter. This is the third, high-eccentricity migration model of HJ formation.

The pericenter of a Hot Jupiter candidate has to be extremely close to the star for tidal forces to be effective \citep[e.g.,][]{Hut1981, Wang2022}. Many authors have proposed the von Zeipel-Kozai-Lidov mechanism  \citep[hereafter ZLK,][]{Zeipel1910, Kozai1962, Lidov1962} or a combination of planet-planet scattering and ZLK mechanism for HJ formation; it leads to an oscillation of inclination and eccentricity and thus to the required close pericenter passage \citep{Fabrycky2007, Nagasawa2008, Naoz2011, Naoz2012, Munoz2016, Anderson2016}. \\
Alternatively, Hot Jupiters can be produced by plain planet-planet scattering or secular chaos within the planetary system \citep{Chatterjee2008, Wu2011, Beauge2012, Hamers2017, Teyssandier2019}. Both scenarios can be realized in a star cluster environment. \\

In the following we will discuss further studies of Hot Jupiter formation in star clusters. They are very diverse with regard to their methodology, the properties of the star cluster, and the architecture of the planetary system. Therefore a quantitative comparison between these studies is difficult.
\citet{Shara2016} used direct $N$-body simulations of the dynamical evolution of Jupiter-mass planets in two-planet systems in a star cluster with 22\,000 members. 
They used special routines in the $N$-body code tailored for hierarchical three-body systems, which means they could not extend their models to planetary systems with more than two planets and a single host star.
\citet{Hamersetal2017} found an effective Hot Jupiter formation probability of about 2\%, in the case where a Jupiter-mass single planet orbits initially at a semi-major axis of 1\,au and the stellar density of the surrounding star cluster is between $10^3$ and $10^6$ particles per pc$^3$.
\citet{Wang2020, Rodet2021} showed with an ensemble of simulations that stellar fly-bys in a two-planet system can dynamically form Hot Jupiters via a von Zeipel-Lidov-Kozai mechanism. These findings are considered valid both for typical open and globular star clusters. \citet{Wang2022} took the possible evolution of three-planet systems after just one stellar fly-by per individual planetary system and showed that secular chaos enhances the Hot Jupiter fraction, especially in an open cluster environment.
The outermost planet in \citet{Wang2020} had Saturn-mass, whereas in \citet{Wang2022} the outermost planet had either Jupiter-mass or Neptune-mass; in both studies the inner planets were Jupiter-mass planets. The outer Jupiter-mass planet was found to be more effective in triggering Hot Jupiter formation. Recently, \citet{Lietal2023} showed that fly-bys can effectively create Hot Jupiters via the von Zeipel-Lidov-Kozai mechanism in a single-planet system in an open star cluster environment. The Hot Jupiter fraction, in \citet{Lietal2023}, depends on the binary fraction of the star cluster, e.g., a HJ probability of 10 \% at a binary fraction of 0.3 and 18\% at a binary fraction of 1.
All previously cited papers, excluding \citet{Shara2016}, used scattering experiments only, where fly-by encounters of stars with planetary systems are followed by a few body integration of the planetary system; the star cluster environment is only taken into account statistically by using a Monte Carlo procedure to pick the encounter parameters for the fly-by.
\\
Here in our paper we will use direct $N$-body simulations of star clusters to deliver the information about encounters for our
planetary systems with 3, 4, or 5 Jupiter-mass planets. We aim to evaluate the possibility of Hot Jupiter formation in star clusters with initially 
high stellar density in the core.
We adopt the computational approach of \cite{Cai2017},(\citeyear{Cai2018}),(\citeyear{Cai2019}), \citet{Dotti2019} and \citet{Stock2020}
as detailed below in Sec.~\ref{LPSsubsect}. In each model, we insert 200 planetary systems into the star cluster with 32 000 and 64 000 members. The approach can reproduce the formation of Hot Jupiters in star clusters and ensures that stellar fly-by parameters are realistic. Furthermore, it considers the star cluster's dynamical evolution and the cumulative effect of multiple stellar fly-bys on the planetary systems. The dynamical evolution of the planetary systems in our star clusters is followed for 100 Myrs; two types of initial planetary systems are investigated: (i) planetary systems with gas giants beyond a semi-major axis of 5\,au and (ii) planetary systems where the innermost planet is at a semi-major axis of 1\,au to account for the possibility of primordial disc-migration or in-situ formation. Moreover, we evaluate whether the formation of young Hot Jupiters with an age of less than 100 Myrs \citep[e.g., 17 Myrs old HIP 67522 b][]{Rizzuto2020} is possible via this channel.  \\
The paper is structured as follows: In Sec.~\ref{High-eccentricity migration} we will briefly summarise the background of the high-eccentricity migration in planetary systems. In Sec.~\ref{Methodology} we will explain in more detail the computational approach. In Sec.~\ref{Initial Configuration} we will give the initial configuration of star clusters and planetary systems. In Sec.~\ref{Results}, \ref{Discussion} and \ref{Conclusion} we will present and discuss the results and conclusions. 

\section{High-eccentricity migration}
\label{High-eccentricity migration}
High-eccentricity migration can dynamically form a Hot Jupiter without in-situ formation or disc migration. The planet's orbit is dynamically excited to high eccentricities by interactions with other planets or stars, while its pericenter approaches the host star. The most important mechanisms that create such high eccentricities are described in Sec.~\ref{High-eccentricity mechanisms}. Tidal forces then shrink the semi-major axis and eccentricity of the planet, forming a tight and circular orbit. This process is complex and not yet well understood, neither qualitatively nor quantitatively \citep[e.g.,][]{Ogilvie2014, Teyssandier2019}. In Sec.~\ref{Tidal migration}, we briefly summarize the tidal model used in this work. 

\subsection{High-eccentricity mechanisms}
\label{High-eccentricity mechanisms}

\subsubsection{Direct scattering}
\label{Direct scattering}
Gravitational interactions with fly-by stars can excite the orbits of the Jupiter-mass planets to high eccentricities \citep[e.g.,][]{Spurzem2009, Dotti2019}. In order to achieve Hot Jupiter formation, the pericenter has to be in the interval of roughly 0.01~-~0.05\,au, where tidal forces are effective but the planet is not yet tidally disrupted (see Sec.~\ref{Tidal migration}). If we consider a Hot Jupiter candidate with a semi-major axis of 5\,au (respectively 1\,au), this yields an eccentricity window of $e \approx 0.990-0.998$ (respectively $e \approx 0.95-0.99$).  \citet{Hamersetal2017} found that only close-orbiting planets in single-planet systems with a semi-major axis of 1\,au and stellar environments with number densities of $n_* \approx 10^4 \rm pc^{-3}$ can be effective for Hot Jupiter formation. This corresponds to the central regions of our chosen clusters (see Sec.~\ref{Star Cluster configuration}). On the other hand, star clusters with lower number density ($n_* < 10^3 \rm pc^{-3}$) or planets beyond a semi-major axis of 5\,au rarely create Hot Jupiters through direct scattering \citep[e.g.,][]{Hamersetal2017, Wang2020, Lietal2023}.

\subsubsection{von Zeipel-Lidov-Kozai mechanism}
The von Zeipel-Lidov-Kozai mechanism describes the oscillation between eccentricities and inclination of one of the components (e.g. a planet) in a binary system due to the presence of an outer companion. At the eccentricity peaks, this can lead to a pericenter of a planet close enough to the host star so that the tidal forces can set in. \\ 
As an example, we consider the simplest case of a planet orbiting one star in a stellar binary system with a small mass in comparison to the stars. If the initial inclination $i_0$ of the planet with respect to the plane of the two stars is about 40° or higher \citep[e.g.,][]{Naoz2016} the inclination and eccentricity can oscillate and the eccentricity can reach values of 
\begin{equation}
    e_{\rm max}=\sqrt{1-\frac{5}{3}\cos^2 i_0}
\end{equation} 
\citep[e.g.,][]{Naoz2016} because $\sqrt{(1-e^2)}\cos i \approx \rm const$. This can yield a pericenter close enough to the host star and form a Hot Jupiter \citep{Nagasawa2008}.
Recently, \citet{Wang2020, Rodet2021} showed that fly-bys can trigger the ZLK mechanism in a system with 2 massive planets in star clusters and the latter creates Hot Jupiters. But also single-planet systems in star clusters can form Hot Jupiters via the ZKL mechanism if the host star acquires a stellar companion \citep{Lietal2023}.  \\
In a hierarchical triple system with an inner binary (e.g., host-star and the inner planet) and an outer binary (e.g., stellar binary or host-star and the outer planet) the timescale for the oscillation can be estimated by:  
\begin{equation}
    t_{\rm ZLK} \sim P_{\rm in} \frac{M_{\rm in}}{M_{\rm out}} \bigg(\frac{a_{\rm out}}{a_{\rm in}}\bigg)^3 (1-e_{\rm out}^2)^{3/2}
\end{equation}
\citep[e.g.,][]{Wang2022} where $M_{\rm in}$ is the mass of the inner binary, $M_{\rm out}$ is the mass of the outer perturber, $a_{\rm in}$ and $a_{\rm out}$  are the mean distance of the inner and outer binary and $P_{\rm in}$ is the orbital period of the inner planet. \\
Additional gravitational influence such as further fly-bys or an additional planet in the planetary system can prevent or end the ZLK mechanism. However, a stellar fly-by can create the right environment by ejecting planets and inclining the orbits of planets even in a coplanar planetary system with $ \geq 3$ Jupiter-like planets \citep[e.g.,][]{Wang2022}.

\subsubsection{Planet-planet scattering}
\label{Planet-planet scattering}
Especially for closely packed planetary systems, planet-planet scattering can create Hot Jupiters \citep[e.g.,][]{Chatterjee2008, Wang2022}. Planets undergo several close encounters, resulting in more eccentric orbits on a typical time scale of a few hundred thousand to Myrs. It has been proposed \citep[e.g.,][]{Wang2020} that the upper limit of eccentricity that can arise from planet-planet scattering of two planets before the planets collide can be estimated by
\begin{equation}
\label{Eq. planet-planet scattering}
    e_{p-p} \approx \frac{\sqrt{G M_p/R_p}}{2\pi a_p/P }
\end{equation}
 where $M_p$ is the mass of the planet, $R_p$ is the radius of the planet, $a_p$ is its semi-major axis and $P$ is its orbital period. Eq.~\ref{Eq. planet-planet scattering} suggests that especially for close-orbiting planets planet-planet scattering is plausible for the formation of Hot Jupiters. A planet of Jupiter's mass and radius at a semi-major axis of approximately 1 au can reach eccentricities close to unity, which enables the possibility of high eccentricity tidal migration \citep{Wang2020}. \\
 In contrast, in planetary systems with three gas giants, other studies, such as \citet{Carrera2019}, did not identify a maximum eccentricity of planet-planet scattering.    

\subsubsection{Secular chaos and angular momentum deficit}
\label{AMD}
In contrast to planet-planet scattering, largely separated planets can exchange significant amounts of angular momentum over timescales ranging from millions to tens of millions of years. This can lead to high eccentricities, eventually resulting in the formation of a Hot Jupiter \citep[e.g.,][]{Wu2011, Teyssandier2019, Wang2022}.
One way to estimate the potential of a planetary system to become unstable is to measure the \textit{angular momentum deficit} \citep[hereafter AMD,][]{Laskar1997, Laskar2000, Laskar2017}. AMD is defined as the amount of lacking angular momentum in comparison to a circular and coplanar orbit with the same semi-major axis. The more eccentric and inclined the orbits of the planets, the higher the chance that the planetary system might become unstable. This instability is a prerequisite for the formation of Hot Jupiters  via secular evolution \citep[e.g.,][]{Wu2011}.\\
Consider a planetary system with a host star of mass $M_0$ and $n$ planets, indexed according to their distance from the host star. The innermost planet has an index of 1. We denote the mass, semi-major axis, inclination, and eccentricity of the planet with index $j$: $M_j$, $a_j$, $i_j$, and $e_j$, respectively, with $M_0\gg M_j$. The angular momentum deficit is then defined \citep[e.g.,][]{Laskar2017} as follows:
\begin{equation}
\label{eqAMD}
    C = \sum_{j=1}^{n}\Lambda_j \bigg(1-\sqrt{1-e_j^2}\cos i_j\bigg)
\end{equation}
where $\Lambda_j=M_j\sqrt{ M_0 a_j \rm{G}}$ is the circular angular momentum of a planet $j$ and G is the gravitational constant. In our simulations, the AMD is due to gravitational interactions with perturbing stars. The total AMD of the planetary system is conserved during secular interactions with other planets. However, within this limitation, eccentricities can shift from outer to inner planets. Therefore, the AMD of a system is crucial in the formation of Hot Jupiters \citep{Wu2011}. It is proposed that a system has to have at least an AMD of
\begin{equation}
\label{eqAMDcrit}
    C_r =\Lambda_1 \bigg[1-0.3 \bigg(\frac{p_{\rm min}}{0.05 \rm{au}}\bigg)^{1/2}\bigg(\frac{a_1}{1 \rm{au}}\bigg)^{-1/2} \cos i_{1,f}\bigg]
\end{equation}
\citep[e.g.,][]{Teyssandier2019} to be able to form a Hot Jupiter.  $p_{\rm min}$ represents the required pericenter distance to the star so that tidal forces become effective and $i_{1,f}$ indicates the final inclination of the inner planet. With a typical value for $p_{\rm min}$ of 0.05\,au \citep[e.g.,][]{Teyssandier2019, Wang2022}, no required final inclination ($i_{1,f}= 0$), and the inner Jupiter-like planet semi-major axis at 5\,au (respectively 1\,au), Eq.~\ref{eqAMDcrit} yields $C_r =  0.87 \Lambda_{\rm 1,5au} (0.7 \Lambda_{\rm 1,1au})$ where $\Lambda_{\rm 1,5au}$ is the angular momentum of the innermost planet at a semi-major axis of 5 au. Because $\Lambda_1 \propto \sqrt{a_1}$, the absolute required amount of AMD for a system with the innermost planet at a semi-major axis of 1\,au is just 36 \% of a planetary system with the innermost planet at a semi-major axis of 5\,au.   \\ 
\cite{Wu2011} have shown that even 3 gas giants with moderate eccentricity might already provide a sufficient amount of AMD. We also measure the AMD of the planetary system and observe the correlation between AMD and the appearance of Hot Jupiters. However, a high amount of AMD should not be seen as a sufficient condition for instability or even Hot Jupiter formation. A system might be AMD-unstable but never destabilizes because the AMD never propagates inwards or leads to destructive effects on the planetary architecture. 

\subsection{Tidal migration}
\label{Tidal migration}
\subsubsection{Constant time-lag model}
After the planet reaches a close pericenter to the host star it will become a Hot Jupiter due to tidal forces. We use the \textit{constant time-lag model} (sometimes also called weak-friction theory) by \citet{Hut1981, Eggleton1998}. Both, the star and the planet exert gravitational forces on each other leading to tidal bulges according to the equilibrium tides theory. However, the amplitude of the tidal bulge is delayed by a constant time-lag $\tau$ relative to the deforming gravitational potential. This misalignment results in energy dissipation.\\ 
In theory, the star exerts its gravitational potential field on the planet leading to tides on the planet and vice-versa. However, on a timescale of 100 Myrs the dissipation within the planet is considered to be dominant \citep[e.g.,][]{Fabrycky2007, Teyssandier2019, Wang2022}. The semi-major axis $a$ and the eccentricity $e$ decrease with time $t$:

\begin{multline}
\label{oshr}
    \frac{da}{dt}=-6\frac{k_2 \tau M_p G}{R_p^3}q(1+q)\bigg(\frac{R_p}{a}\bigg)^8\frac{a}{(1-e^2)^{\frac{15}{2}}}  \\\ \times \bigg[f_1(e^2)-(1-e^2)^{\frac{3}{2}}f_2(e^2)\frac{\Omega_p}{n}\bigg]\\
\end{multline}

\begin{multline}
\label{eshr}
   \frac{de}{dt}=-27\frac{k_2 \tau M_p G}{R_p^3}q(1+q)\bigg(\frac{R_p}{a}\bigg)^8\frac{e}{(1-e^2)^{\frac{13}{2}}} \\\ \times
   \bigg[f_3(e^2)-\frac{11}{18}(1-e^2)^{\frac{3}{2}}f_4(e^2)\frac{\Omega_p}{n}\bigg]\\
\end{multline}
%\begin{equation}
%\label{Omshr}
%    \frac{d \Omega_p}{dt}=3\frac{k_2}{T}\frac{q^2}{r_g^2}\bigg(\frac{R_p}{a}\bigg)^6\frac{n}{(1-e^2)^{6}}\bigg[f_2(e^2)-(1-e^2)^{\frac{3}{2}}f_5(e^2)\frac{\Omega_p}{n}\bigg]\\
%\end{equation}
with the mass relation $q=\frac{M_*}{M_p}$, the planetary spin $\Omega_p$ and the mean motion $n=\sqrt{GM_*/a^3}$. The Love number of second degree $k_2$ characterizes the deforming potential of the planet \citep{Love1911}. 
% radius of gyration $r_g$ defined by $$r_g = \sqrt{\frac{I}{M_p R_p^2}}$$ with the moment of inertia of the primary $I$ and the gravitational constant G.
The $f$ polynomials are defined as 
%, T= \frac{R_1^3}{M_p G \tau}$$\\
\begin{equation*}
f_1(e^2)=1+\frac{31}{2}e^2+\frac{255}{8}e^4+\frac{185}{16}e^6+\frac{25}{64}e^8
\end{equation*}
%$$f_3(e^2)=1+\frac{15}{4}e^2+\frac{15}{8}e^4+\frac{5}{64}e^6$$
%$$f_4(e^2)=1+\frac{3}{2}e^2+\frac{1}{8}e^4$$
\begin{equation*}
f_2(e^2)=1+\frac{15}{2}e^2+\frac{45}{8}e^4+\frac{5}{16}e^6
\end{equation*}
\begin{equation*}
f_3(e^2)=1+\frac{15}{4}e^2+\frac{15}{8}e^4+\frac{5}{64}e^6
\end{equation*}
\begin{equation*}
f_4(e^2)=1+\frac{3}{2}e^2+\frac{1}{8}e^4
\end{equation*}
\begin{equation*}
f_5(e^2)=1+3e^2+\frac{3}{8}e^4
\end{equation*}
 For $t \rightarrow \infty$ the planet reaches a stable orbit at roughly 
\begin{equation}
    a_{\rm final}= a_{\rm initial}(1-e_{\rm initial}^2)  .
\end{equation}
However, a spin evolution model is not yet implemented in our simulations and we use $\Omega_p=0$ in Eq.~\ref{oshr} and Eq.~\ref{eshr}. Consequently, our Hot Jupiters never reach a stable orbit. Furthermore, the spin can counteract the shrinking of the semi-major axis (see Eq.~\ref{oshr}). Without spin implementation, the eccentricity and semi-major axis decreases slightly faster than simulations with spin implementation. Thus, we choose $k_{2}= 0.38$ \citep{Gavrilov1977} and a rather low value of the observational poorly constrained parameter $\tau=0.11$s. This yields a lower $k_{2} \tau$ factor than for other authors \citep[e.g., $k_{2}=0.25, \tau=0.66$s][]{Hamers2017, Wang2020, Wang2022}{}{} who included the planetary spin evolution.  
\subsubsection{Tidal disruption}
\label{Tidal disruption}
If a planet's pericenter is too close to the host star the planet will be tidally disrupted due to the gravitational gradient. This can be expressed by
\begin{equation}
    r_{\rm dis}=\eta R_p \bigg(\frac{M_*}{M_p}\bigg)^{1/3}
\end{equation}
\citep[e.g.,][]{Guillochon2011} with a value of $\eta$ ranging between 1.67 \citep{Naoz2012} and 2.7 \citep{Guillochon2011}. The higher values of $\eta$ take into account the tidal disruption of the planet over many orbital periods. In our work, we consider any planet with a pericenter below $r_{\rm{dis}}(\eta=2)$ as destroyed and remove it from our simulation.\\
Because the spin is not yet implemented in our simulation all Hot Jupiters will eventually undergo tidal disruption. However, because the pericenter remains roughly constant during the circularisation of the orbit by tidal forces this happens after the Hot Jupiter has already been formed. 

\section{Methodology}
\label{Methodology}
The combined simulation of star clusters and planetary systems is challenging for several reasons. The planetary systems evolve through secular evolution provided by mutual gravitational interactions between planets and external perturbation of passing stars of the star cluster, whereas the star cluster evolves through two-body relaxation and few-body encounters. A direct $N$-body simulation is therefore not useful. The dynamical evolution of a star cluster as well as some planetary systems after an encounter also exhibits deterministic chaos.  To accurately display the effects on the internal evolution of the planetary system, the star cluster and the planetary systems are simulated with different codes: \nb{} for the star clusters and ``\texttt{Lonely Planets Scheme}'' (\texttt{LPS}) for the planetary systems. \\
We use the same approach as \citet{Cai2017,Cai2018,Cai2019, Dotti2019, Stock2020}. It is assumed that the orbital evolution of the planets is affected by the neighboring stars but planets do not affect the stellar kinematics.

\subsection{\nb{}}
 The star cluster simulation was carried out by \nb{} \citep{Wangetal2015, Wang2016, Kamlah2022} which integrates the motion of its constituents via 4th-order Hermite Scheme. The code also makes use of the graphical processing units (GPU) as well as parallelization of tasks through a message interference (MPI), resulting in significant speed-up \citep{spurzem2008, nitadori2012}. The output of fundamental parameters is stored at a high time resolution using the "block time storage" (BTS) scheme \citep{Cai2015}. This allows us to use those data to accurately reconstruct the stellar encounters on the planetary system.  \\
 
\subsection{LPS}
\label{LPSsubsect}
We use \texttt{LPS} \citep{Cai2017,Cai2018,Cai2019, Dotti2019, Stock2020} to distribute 200 identical planetary systems around the  host-star in the star cluster whose
masses are closest to 1~$M_\odot$. All host stars are in a mass range of $[0.975-1.025]M_\odot$. We then simulate the dynamical evolution of the planetary system with the perturber data from a pre-recorded \textsc{Nbody6++GPU} star cluster simulation, in which stellar encounters can be resolved in high temporal resolution thanks to the block time step format \citep{Cai2015}. At each of its time steps, \texttt{LPS} selects the five nearest stars on a host star and their gravitational impact on the planetary system is taken into account. Their identity and positions are constantly updated when the perturbing stars change their relative position to the host star \citep{Cai2017}. For intermediate states between two adjacent time steps of the star cluster data, interpolation is applied to calculate the positions of the five closest perturbers  \citep[for a detailed description, see][]{Cai2017, Cai2019}. We neglect the gravitational effect of further stars in the cluster and distant, very massive objects, such as black holes.  
\texttt{LPS} is based on the \texttt{AMUSE} framework \citep{PortegiesZwart2011, PortegiesZwart2018} and it uses \texttt{REBOUND} \citep{Rein2012} and \texttt{REBOUNDx} \citep[][]{Tamayo2020} for the integration. The 15th-order integrator \texttt{IAS15} \citep{Rein2015},  with adaptive integration step size, is used to integrate the planetary system using the stellar dynamics data from the \textsc{Nbody6++GPU} simulation.  Additionally, we implemented the effects of tidal dissipation of \texttt{REBOUNDx} \citep{Baronett2022} in \texttt{LPS} so that we can observe the effect of the tidal forces between the star and the planets.\\
 Our approach allows us to set up a stable planetary system, which is perturbed by realistic encounters of passing stars in the star cluster. Contrary to the similarly motivated paper by \cite{Wang2020, Rodet2021, Wang2022}, which had a system of two and three planets disturbed by a set-up flyby, we can simulate close encounters between the star cluster members.  It ensures that the parameters of fly-bys, like frequency of close encounters as well as mass, velocity, or distance of the perturbing star towards the planetary systems are realistic.
 The cumulative effect of many consecutive stellar encounters, including their possible destructive impacts on the planetary system, are taken into account. Because we use a direct $N$-body integration, we can simulate planet-planet scattering and secular chaos of the planetary system very accurately even though this is computationally expensive and limits the sample size.  \\
\\
\section{Initial Configurations}
\label{Initial Configuration}
\subsection{Star Clusters}
\label{Star Cluster configuration}
\begin{table} 
\centering
\begin{tabular}[H]{l|c|c}
 \hline
 Star Cluster  & C32k & C64k \\
 \hline
$N$ & 32 000 & 64 000 \\
$M_{\rm tot}$ &  16 302 $\rm M_\odot$ & 32 619 $\rm M_\odot$ \\
$r_{\rm hm}$ & 0.78 pc & 0.78 pc \\
$r_{\rm tid}$ & 35.84 pc & 45.16 pc \\
$\rho_{\rm central}$ & 13 852 $\rm M_\odot \rm{pc}^{-3}$ &  25 153  $\rm M_\odot \rm{pc}^{-3}$\\
 \hline
\end{tabular}
\caption[Tabelle]{Initial configuration of the star clusters where $N$ is the number of members, $M_{\rm tot}$ the total mass, $r_{\rm hm}$ the half mass radius, $r_{\rm tid}$ the tidal radius and $\rho_{\rm central}$ the central density.}
\label{StarClusters config}
\end{table}
\begin{figure}
\centering
\includegraphics[width=8cm]{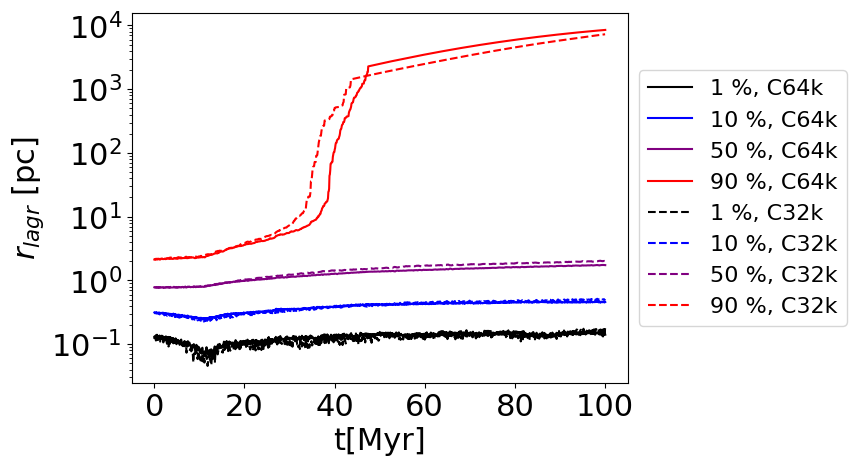}
\caption[Figure]{Lagrange radii $r_{\rm lagr}$ of 1, 10, 50, 90 \% of the C32k and the 64k star cluster. The central regions undergo a core collapse within the first 20 Myrs. Thereafter all regions of the star cluster are expanding. The C32k star cluster expands faster increasing the density difference between the star clusters. }
\label{Lagrradii}
\end{figure}

In this work, we use star clusters with 32 000 and 64 000 members (here called C32k and C64k) similar to \citet{Stock2020}. A summary of the most important initial properties of the star clusters can be found in Tab.~\ref{StarClusters config}.  Both models of the star clusters were set up according to the initial mass function of \citet{Kroupa2001} with a mass range of $0.08-100 \rm M_\odot$ and an average mass of $0.51 \rm M_\odot$. The evolution of mass and radii of the stars follows \citet{Hurley2005} i.e., evolution level B as defined in \citep{Kamlahetal2022}. The initial velocities and positions were drawn from the \citet{Plummer1911} model with a virial radius of one pc, which has a half mass radius of $r_{\text{hm}}=0.78$ pc and a central density of $13 852\,(25153)\,\rm M_\odot \rm{pc}^{-3}$ for 32k (64k) particles, respectively. We use $r_{\rm tid}=R_G (M_{\rm cl}/ M_G)^{1/3}$ to calculate the tidal radius, where $R_G$ and $M_G$ are the distance to the Galactic center and the mass of the Galaxy contained inside $R_G$. $M_{\rm cl}$ is the star cluster mass. We assume that our star clusters orbit the galactic center as the solar neighborhood and are affected by the tidal forces of the galaxy in the same way \citep{Heisler1986}. 
For simplicity, primordial binary systems and primordial mass segregation \citep[i.e., heavier stars are located preferentially in the center, see][]{fischer1995} are not included. Based on the assumption that the average mass of the encountering star is $0.5\,\rm M_\odot$ on the planetary system's host star with one solar mass at a periastron distance of $r_{\rm min}=1000$ au, the C64k has an average encounter timescale of roughly $1.1$ Myrs \citep[see equation 3 in][]{Malmberg2007}. In our case, the total mass of the stars involved in the encounter is $m_t=1.5\,\rm M_\odot$ \citep[cf.][]{Stock2020} instead of the total mass value of $m_t=1\,\rm M_\odot$ for the stars engaged in the encounter as originally done in \citet{Malmberg2007}. \\ 
The Lagrangian radii yield the fraction of the total mass of the star cluster within a certain radius. In Fig.~\ref{Lagrradii} the Lagrangian radii of 1, 10, 50, and 90 \% of the star clusters are depicted describing the dynamical evolution.  After a short phase of core collapses the cluster expands on a fast scale and the central density decreases. As a result, the star cluster will lose stars over time \citep{Ernst2008} even though in our simulation we do not remove ejected stars. Thus at the start of the simulation, the host stars with a planetary system could be in a rather dense environment but typically will be in a much less dense stellar neighborhood in the later stages of the simulation due to the onset of mass segregation and dissolving of the star cluster. For the majority of these stars, the stellar density is for most of the time in the order of up to a few hundred $\rm M_\odot \rm{pc}^{-3}$ \citep{Stock2020}. We expect most strong encounters to happen in the first 100 Myrs. \\
The dynamical evolution of both star cluster models is comparable. Within the first 20 Myrs, both experience a brief phase of core collapse in their central regions. Throughout the simulation period, the inner most considered region (Lagrangian radius 1\%) maintains a relatively similar density for both star clusters. \\ 
Overall, both star clusters expand during the simulation. The initial half-mass radius is identical for both star cluster models. The C64k star cluster has about twice the mass of C32k and has therefore a higher initial density. The rate of expansion is greater in C32k than in C64k, resulting in an amplified density difference between the two star clusters for most stars. As mentioned, Sun-like stars are very unlikely to be found in the central regions in the later stages due to mass segregation. Overall, most of the Sun-like stars will undergo significantly more close stellar encounters in the C64k star cluster than in the C32k model, especially in the later stages of the simulation. \\
\citet{Lietal2023} recently found a correlation between the rate of binaries in open star clusters and the formation of Hot Jupiters. As primordial binaries are not included only a small number of binaries arise during the simulation time. Thus, Hot Jupiter formation involving a stable stellar binary is unlikely to take place in our simulation. 

\subsection{Planetary Systems}
\label{Planetary System}
In the simulation, we use 200 host stars in each star cluster model (C32k and C64k) with the mass and radius of the Sun, as Hot Jupiters are mostly found around main-sequence stars  \citep[e.g.,][]{Dawson2018}. 
The choice of the host star's initial position in the star cluster is arbitrary, i.e., we give no preference towards a certain region of the star cluster in our simulations. Around each of those 200 host stars, we set up the same planetary system architecture. For both star clusters, we simulate three times the dynamical behavior of these planetary systems for 100 Myrs and use in each simulation a different planetary architecture. \\  
All planets have the mass and radius of Solar System's Jupiter. Recent simulations suggest that planetary systems with Jupiter mass planets are favorable to dynamically create Hot Jupiters \citep[e.g.,][]{Anderson2016, Wang2022}. The goal of this work is to see how Hot Jupiters are formed in a multi-planetary system. Therefore we use systems with either 3, 4, or 5 Jupiter-like planets. The initial semi-major axes are listed in Tab.~\ref{PlanetsAU}. The closer the semi-major axis of the innermost planet, the higher the likelihood of Hot Jupiter formation. Without primordial disc-migration or in-situ formation, the semi-major axis of the closest Jupiter-like gas giant in a planetary system is widely considered to be at 5\,au \citep[e.g.,][]{Wang2020, Rodet2021, Wang2022, Lietal2023}. However, we also want to investigate the possible combined effect of primordial disc-migration or in-situ formation and high-eccentricity migration. Therefore, in the 5-planet system (hereinafter called 5PL1a) the closest orbiting planet is at 1\,au whereas it is at 5\,au for the 4 and 3-planet system (hereinafter called 4PL5a and 3PL5a). \\
Generally, planets orbiting further outwards are more affected by a fly-by. On the other hand, close-packed systems are more likely to exhibit planet-planet scattering and might propagate angular momentum among each other easier increasing the chances for a Hot Jupiter to form. Furthermore, the further out a planet, the more likely that planet is ejected (see Sec.~\ref{Survival}). We choose the distance between the Jupiter-mass planets to be roughly equal to the distances of the gas giants in the Solar system with the furthest orbiting planet at 30\,au in the 4- and 5-planet system and 20\,au in the 3-planet system \citep[cf.][]{Stock2020}. As giant plants are expected to have low eccentricities and inclinations after the protoplanetary disc has dissolved \citep[e.g.,][]{Kley2012}, we give the planets no initial eccentricity or inclination. This results in an initial AMD of zero. We do not take into account any stellar or planetary spin. Without external perturbation, the system is stable over the simulation time of 100 Myrs and shows no signs of any resonances. Planets with an eccentricity of $e>0.999$ or below the tidal radius (see Sec.~\ref{Tidal disruption}) are considered to be ejected or respectively destroyed and are excluded from the simulation.  
\begin{table} 
\centering
\begin{tabular}[H]{l|c|c|c|c|c|c|r}
 \hline
  Title & P1 [au] & P2[au] & P3[au] & P4[au] & P5[au] & $e$ & $i$ [°]\\
 \hline
5PL1a & 1 & 5 & 10 & 20 & 30 & 0 & 0\\
4PL5a  & - & 5 & 10 & 20 & 30 & 0 & 0\\
3PL5a  & - & 5 & 10 & 20 & - & 0 & 0 \\
 \hline
\end{tabular}
\caption[Tabelle]{Initial orbital parameters of the planets. The initial eccentricity $e$ and inclination $i$ is 0. All planets have the mass of $1M_{\rm Jup}$. We name our planetary models according to their number of planets and the position of the innermost planet.}
\label{PlanetsAU}
\end{table}
\\      
\section{Results}
\label{Results}

\subsection{Terminology}
Generally, a Hot Jupiter formation via secular high-eccentricity migration evolves through three phases \citep[e.g.,][]{Wu2011}:  (i) the system has enough AMD to form a Hot Jupiter. In our simulations, this happens due to close stellar encounters. Then, (ii) when a planet reaches high eccentricities, tidal forces start to become effective. Finally (iii), if tidal forces continue to affect the orbit of the planet for a sufficient duration, an actual Hot Jupiter is formed. Depending on the pericenter distance and the strength of tidal forces, the circularization time can range from a few tens of thousands of years to billion of years \citep{Wang2022}. Before we go through the results, we define the acronyms used in this section. 
\begin{itemize}
\item \textbf{Sufficient AMD of planetary systems} (SAMD): Planetary systems that have sufficient angular momentum deficit ($C > C_r$) to form a Hot Jupiter according to Sec.~\ref{AMD} for at least 100 kyrs.  \\ 
\item \textbf{Hot Jupiter candidate} (HJC): Because the circularisation time is usually several times longer than our simulation time of 100 Myrs, we adopt the semilatus-rectum criterion of \citet{Hamers2017}. A planet is classified as a Hot Jupiter candidate if  $a_p(1-e_p^2)<0.091$\,au for a duration of at least 100 kyr. This is equivalent to a Hot Jupiter having a final orbital period of 10 days and requires a pericenter $<$~0.05\,au. We extend the period for which the semilatus-rectum criterion must be met to   100 kyrs to avoid incorrectly classifying planets as Hot Jupiter candidates which have the pericenter close to the host star for a very short time. This is for example the case if planets are ejected or tidally disrupted. \\
\item \textbf{Young Hot Jupiter} (YHJ): A planet which has developed into a Hot Jupiter (orbital period <10\,days) within our simulation time of 100 Myrs. All YHJs are also categorized as HJCs. \\
\end{itemize}
   
\subsection{General dynamic and survival fraction}
\label{Survival}

\begin{figure}
\centering
\includegraphics[width=8cm]{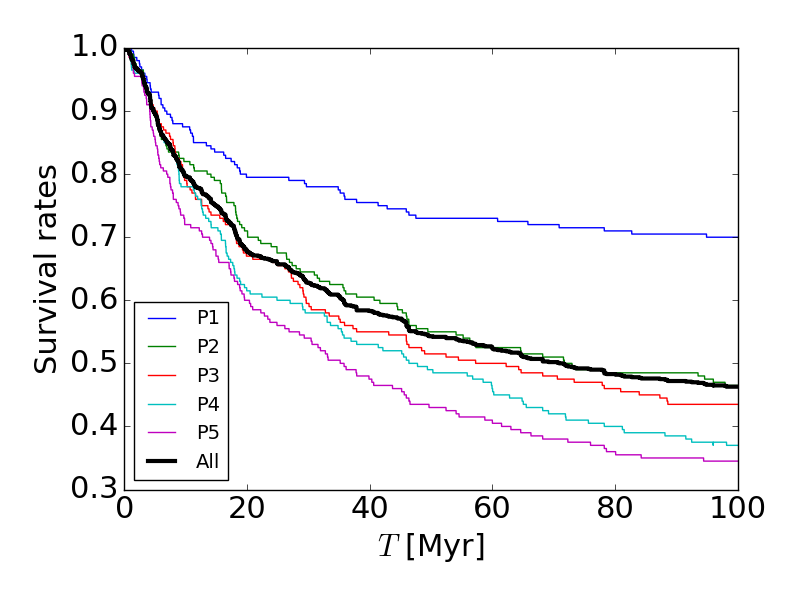}
\caption[Figure]{Survival fraction of planets of the 5PL1a model in the C64k star cluster. The colored lines represent the survival rate of a planet at a particular initial semi-major axis whereas the black line represents the survival rate of all planets. After 100 Myrs more than half of all planets are ejected.}
\label{Suvival_fig}
\end{figure}

\begin{table} 
\centering
\begin{tabular}[H]{l|c|c|c|c|c|c|r}
 \hline
  Star Cluster & model  & P1 & P2 & P3 & P4 & P5 \\
 \hline
C64k & 5PL1a & 70\% & 46.5\% & 43.5\% & 36.5\% & 34.5\% \\
C64k & 4PL5a  & - & 55\% & 46.5\% & 36.5\% & 39.5\% \\
C64k & 3PL5a  & - & 70\% & 61\% & 53\% & - \\
C32k & 5PL1a & 71.5\% & 66\% & 58.5\% & 55\% & 54.5\% \\
C32k & 4PL5a  & - & 70\% & 63.5\% & 55.5\% & 53\% \\
C32k & 3PL5a  & - & 80\% & 72.5\% & 68.5\% & -  \\
 \hline
\end{tabular}
\caption[Tabelle]{Percentage of planets which have survived in C32k and C64k in each model after 100 Myrs. The initial semi-major axis of the planets P1-5 are 1, 5, 10, 20 and 30\,au.}
\label{TabPlanetsSurv}
\end{table}

The general dynamic and survival fraction of planets in star clusters are only briefly summarised in this section. For a more profound analysis of planetary systems in star clusters see \cite{Cai2017},(\citeyear{Cai2018}),(\citeyear{Cai2019}), \citet{Dotti2019} and \citet{Stock2020}. \\

\subsubsection{Survival Rate}
 We consider planets as "survived" if they are not tidally disrupted or ejected, i.e., if $e<0.999$ for 100 Myr \citep{Cai2019, Dotti2019}. Planets can either be ejected directly or indirectly by the close encounter (here defined as encounter distance $r <1000\,\rm au$). Indirect ejection can happen millions of years after the fly-by due to internal planetary dynamics \citep[i.e., secular dynamics, see, e.g.,][]{pu2018}. In this process, host stars may lose all or just some of their planets. We refer to a system as "destroyed" if all planets are ejected.  \\
 According to our results, the majority of planets are ejected from the planetary system and not tidally destructed. Only a low fraction (i.e., approximately 6.5\%) of non-surviving planets experience tidal disruption or collision with the star. However, both tidal disruption and stellar collision account for approximately half of the non-surviving P1 planets in the 5PL1a model for both star clusters.
 An overview of the survival fraction of planets after 100 Myrs can be found in Tab.~\ref{TabPlanetsSurv}. The survival fraction of the 5PL1a as a function of time of the C64k is shown in Fig.~\ref{Suvival_fig}. \\
 The survival rate of the planets in the 5PL1a model in C64k drops sharply in the first 20 Myrs. Overall, by the end of the simulation time of 100 Myrs, more than half of all planets (i.e., 54 \%) are ejected. The surviving percentage of the planets in the 5PL1a model after 100 Myrs are 70\%, 46.5\%, 43.5\%, 36.5\%, and 34.5\%. Generally, the further out the planet is initially, the higher the chances that it is ejected \citep[e.g.,][]{fujii2019}. The survival fraction of the inner planet is significantly higher than the four outer planets. However, P4 and P5 have roughly the same chance of survival in both star clusters. \\
 The results indicate that the more planets are in a system, the lower the survival fraction for the planets which are present in all three planetary models, i.e., P2, P3, and P4. Therefore, the survival fraction of planets in the 3PL5a model is higher than their equivalents in the 5PL1a model.  \\
As mentioned in Sec.~\ref{Star Cluster configuration}, the C32k is generally less dense than the C64k and fewer close encounters take place. In the C32k model, more planets survive, in any of the planetary models. Overall, in the 5PL1a model in the C32k star cluster, 60\% of planets survived. The surviving fraction of the planets in the 5PL1a model after 100 Myrs are 71.5\%, 66\%, 58.5\%, 55.5\%, and 54.5\%. Remarkably, the inner planet P1 at a semi-major axis of 1\,au has almost the same survival rate as in the C64k. Additionally, one notices much less discrepancy in the survival rate of the inner planet and four outer planets in comparison with the C64k.  \\

\begin{figure}
\centering
\includegraphics[width=8cm]{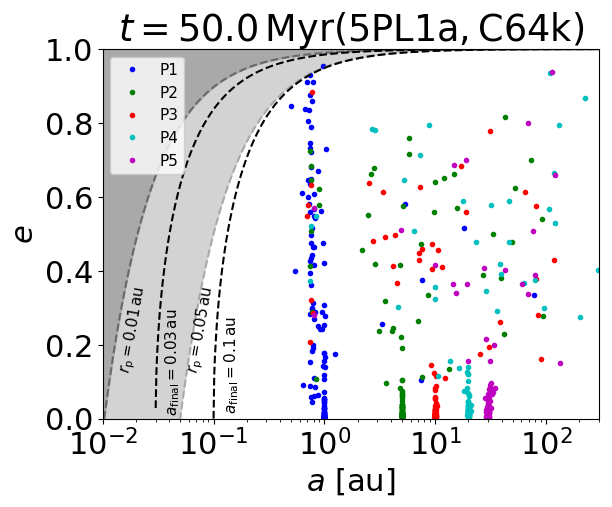}
\includegraphics[width=8cm]{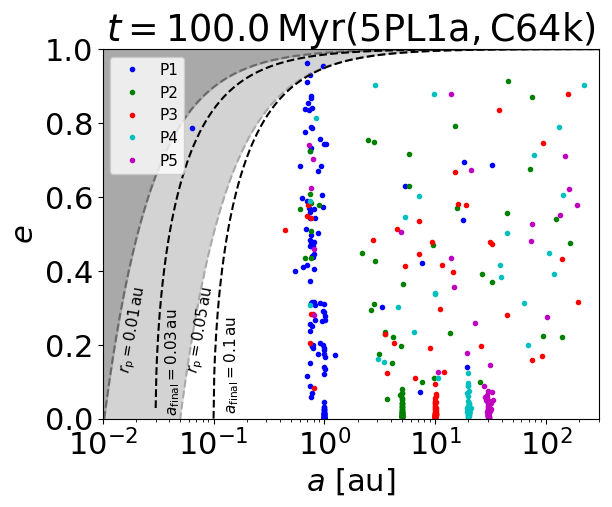}
\caption[Figure]{The semi-major axis-eccentricity parameter space of the 200 5-planet systems (5PL1a model) in the C64k star cluster after 50.0 Myrs and 100 Myrs. The color of the dots indicates their initial position. The dark gray area marks the region of the $a-e$ space where planets could be tidally disrupted ($<0.01$~au), whereas the light gray area represents the region where tidal forces are effective ($<0.05$~au). The dashed lines represent the projected shrinking of the semi-major axis and eccentricity due to tidal forces with constant angular momentum (approximately following the constant time-lag theory, see Sec.~\ref{Tidal migration}) of a planet with a final semi-major axis of 0.03 and 0.1\,au. \\
At both particular instances,  a few planets are affected by tidal forces and can become Hot Jupiters \citep[e.g.,][]{Wu2011}. At 100 Myrs one planet (\#~119) has become an eccentric} young Hot Jupiter.
\label{a-e plot}

\end{figure}

\subsubsection{Dynamics of planetary systems}
 The chosen 200 Sun-like stars are the same for every model in each star cluster. Nevertheless, each chosen planetary architecture would react differently to the same stellar encounter.  \\ 
Fig.~\ref{a-e plot} depicts the $a-e$ space of all bounded planets to their host star, in the C64k with the 5PL1a model at 50 and 100 Myrs. The effect of stellar fly-bys on planetary systems is complex. We limit ourselves to aspects of planetary dynamics in connection with Hot Jupiter formation. \\
As expected, the C64k star cluster destabilizes more planetary systems than the C32k star cluster. Stellar encounters directly or subsequent planet-planet scattering can move planets inwards or outwards. However, even close fly-bys might also not have any destabilizing effect on the planetary system at all. Typical values for the maximum inward migration are approximately 0.7\,au and 3\,au \citep[consistent with][]{Shara2016} for the initial innermost planet at a semi-major axis of 1\,au and 5\,au.  Generally, outer planets are more affected by stellar fly-bys than inner planets. Outer planets, in particular, often change their positional order relative to the host star. Nevertheless, also the innermost planet can alter its position relative to the host star with one of the outer planets. Consequently, the planet initially identified as the innermost may not maintain that position by the end of the simulation. \\
Furthermore, stellar encounters and subsequent planet-planet scattering can heavily perturb the orbital elements of a planetary system. The secular dynamics of the planetary system can further change the planet's orbital elements on timescales from a few thousand years to tens of million years. If the planet reaches a sufficiently close pericenter to the star and if tidal forces are effective (i.e., pericenter $< 0.05$ au), the planet tidally migrates inwards and can become a (young) HJ. \\ 
 We observe several dynamical outcomes in planetary systems affected by close encounters.  In planetary systems with a HJC, on average, one strong encounter followed by violent planet-planet scattering changes the order of the planets. Commonly in this process, at least one planet is ejected. On the other hand, the planet that transforms into a Hot Jupiter candidate undergoes dynamical scattering. This scattering brings an orbit closer to the star than the initial innermost planet. Then, secular evolution within the planetary systems or resonances between two planets, often combined with the impact of additional stellar encounters lead to the required high-eccentricity values. \\
 We provide 2 examples of YHJ formation in Sec.~\ref{Examples for YHJ formation} and a HJC which will not become an actual Hot Jupiter in Fig.~\ref{} in Sec.~\ref{Constraints}. 

\subsection{AMD and Hot Jupiter candidates statistics}

\begin{table} 
\centering
\begin{tabular}[H]{l|c|c|c|c|c|r}
 \hline
  Star Cluster & model & YHJ & HJC & SAMD \\
 \hline
C64k  & 5PL1a  & 0.5\% & 15 \% & 69\% \\
C64k  & 4PL5a  &  0\% & 1.5 \% & 66\% \\
C64k & 3PL5a  &  0\%  & 2 \% & 44\% \\
C32k & 5PL1a  &  0.5 \% & 15 \% & 48 \%\\
C32k  & 4PL5a  &  0\% & 4.5\% & 50 \%\\
C32k & 3PL5a  &  0\% & 1.5\% & 25.5 \% \\
 \hline
\end{tabular}
\caption[Tabelle]{Percentage of simulated planetary systems with a young Hot Jupiter, a Hot Jupiter candidate, or sufficient Angular momentum deficit (SAMD) after 100 Myrs. YHJ are also listed as HJC. }
\label{FraHJ}
\end{table}

The results are summarized in Tab.~\ref{FraHJ}. It shows the percentage of planetary systems with SAMD, systems with at least one HJC, or systems with at least one YHJ. The relation of planetary architecture, star cluster, and AMD is described in detail in section \ref{Sufficient Angular momentum deficit}. The results on Hot Jupiter candidates and young Hot Jupiters are outlined in \ref{Hot Jupiter candidates} and \ref{young Hot Jupiters}.  
\subsubsection{Sufficient angular momentum deficit}
\label{Sufficient Angular momentum deficit}
All planets have been initialized with circular and coplanar orbits. Thus, the initial AMD is 0. Stellar flybys can induce non-zero AMDs. Our results indicate that the fraction of planetary systems with sufficient AMD ($C > C_r$, see Sec.~\ref{AMD}) increases with higher densities and the presence of an additional outer planet.\\
In the C64k, $66-69\%$ of simulated planetary systems with an outer planet at 30\,au (5PL1a, 4PL5a) fulfilled the criterion. In the C32k model, the fraction is roughly 25-40\% lower. The fraction drops more than 30 \% in both star clusters if the outer planet is removed (3PL5a). In C32k, only 25.5\% of planetary systems have SAMD in the 3PL5a model. \\
With the innermost planet at 1 au, 5PL1a requires only 36\% of the absolute AMD amount for secular HJ formation compared to the 3/4PL5a planetary architectures, where the innermost planet is at 5 au (see Sec.~\ref{AMD}). Nevertheless, the impact on the fraction of systems with sufficient AMD is marginal. 5PL1a has in both star clusters roughly the same number of AMD-sufficient systems as 4PL5a (i.e., 69\% to 66\% and 48\% to 50\%). \\ 
AMD is regarded as a required criterion for the secular formation of Hot Jupiter via high-eccentricity migration \citep[e.g][]{Wu2011, Teyssandier2019, Wang2022}. Only a tiny fraction ($2-30$\%) of systems with sufficient AMD turn actually to systems with a HJC. The number is especially low in models with the inner planet at 5\,au. A large fraction of SAMD systems does not seem to correlate with a high fraction of systems with a HJC. For example, 4PL5a in C64k has a significantly higher percentage of SAMD systems than 4PL5a in the C32k (66\% to 48\%). Yet, 4PL5a in the C32k has more planetary systems with a HJC (4.5\% to 1.5\%).

\subsubsection{Hot Jupiter candidates}
\label{Hot Jupiter candidates}
We note in all models that the HJC is not necessarily the initially innermost planet. Planets further out can change positions with inner planets and become HJCs. In the 5PL1a model, where the innermost planet is positioned with a semi-major axis of 1 au, 68\% of the HJCs originate from the initially innermost planet. In contrast, for the 3/4PL5a model, where the innermost planet is at a semi-major axis of 5 au, the percentage is 42\%. This indicates that a planet at a semi-major axis of 5 au is more susceptible to position changes with outer planets compared to a planet at a semi-major axis of 1 au. \\
An additional close-orbiting planet drastically increases the likelihood that a HJC is created. 5PL1a, with an innermost planet at a semi-major axis of 1\,au, produces a high percentage of all planetary systems in both star cluster models with HJC (i.e., $15\%$). Both star clusters dynamically create the same amount of planetary systems with a HJC. \\
4PL5a and 3PL5a have both the innermost planet at a semi-major axis of 5\,au. A significantly higher amount of angular momentum transfer is needed to reach a pericenter below $0.05$\,au (see Sec.~\ref{AMD}). Still, in each of the four models with the innermost planet at a semi-major axis of 5\,au, we found at least three systems ($1.5\%$ of all planetary systems) with a planet that fulfilled the criterion for a Hot Jupiter candidate. The fraction in 4PL5a in the C32k star cluster even reaches $4.5\%$. \\ 
The influence of an additional outer planet seems to be dependent on the star cluster. 4PL5a, in contrast to 3PL5a, has an additional outer planet at 30\,au. In the C64k star cluster, the HJC percentage is at 2\% slightly higher in the 3PL5a model than in the 4PL5a model at 1.5\%. Whereas inserted in the C32k star cluster, 4PL5a shows a significant increase of systems with Hot Jupiter candidates to 4.5\%. The number of planetary systems with HJC is here three times higher than in 3PL5a. \\
A suitable explanation could be: The C64k star cluster has a high frequency of close stellar encounters to introduce sufficient perturbation to the planetary systems even without an additional planet at 30\,au. On the downside, the C64k star cluster also ejects more planets or destroys entirely the system. This prevents the creation of HJCs. The additional outer planet is not a significant advantage in this case. In contrast, the C32k star cluster is less dense and has therefore less strong encounters. There are fewer ejections of planets (roughly 29\%, see Tab.~\ref{TabPlanetsSurv}) or complete destruction of the planetary system. On the other hand, the introduction of enough AMD into the system becomes critical. In this case, the additional outer planet is a significant advantage since outer planets are easier perturbed by stellar fly-bys. \\

\subsection{Young Hot Jupiters}
 \label{young Hot Jupiters}
 \subsubsection{Statistics}
 Young Hot Jupiters within 100 Myrs are only found in the 5PL1a system. No YHJ was found in the 4PL5a and 3PL5a systems with the inner planet set up at a semi-major axis of 5\,au. We believe that some suitable explanations would be: 
 \begin{itemize}
     \item To find a young Hot Jupiter, the circularisation time must be short ($t_{\rm circ} < 100$ Myrs). An even closer pericenter is needed \citep[$<0.025$~au, e.g.,][]{Wang2022}. A close orbiting planet needs less extreme eccentricity values to reach that pericenter distance than further outwards orbiting planets. Therefore, systems with close orbiting planets (like 5PL1a) are preferred. 
     \item The already close orbiting planet is better protected from destructive encounters (see Sec.~\ref{Survival}). 
     \item More planets might transfer perturbations better. Systems with many planets could therefore be preferred. 
 \end{itemize}
 We found in total two YHJ, thereof one in C64k and one in C32k. Therefore, both star clusters seem to form systems with a YHJ at a percentage of roughly 0.5\% of all simulated systems, even though the sample size is very low. This is consistent with the HJC fractions being the same for the 5PL1a model in C64k and C32k. \\
 We note that the circularisation time can even be much lower than 100 Myrs, i.e., a few Myrs (see Fig.~\ref{p_sys_119} and Fig.~\ref{p_sys_37}). Furthermore, the two YHJ have high inclinations, i.e., $i = 57^{\circ}$ and $i = 63 ^{\circ}$ at the moment the planets are classified as YHJs.\\
 So far, there are a few young close transiting planets found  \citep[e.g., HIP 67522 b, DS Tuc Ab, TOI 942 b, V1298 Tau c, AU Mic b][]{Rizzuto2020, Heitzmann2021}. They are seen as evidence for the in-situ or disc-migration formation of Hot Jupiter. Our work confirms that high-eccentricity migration beyond a semi-major axis of 5\,au is unlikely to account for those planets. However, our results suggest that a combination of in-situ or disc-migration and high-eccentricity migration can create young Hot Jupiters. The planet is formed in-situ or undergoes disc migration to a semi-major axis of less than 5\,au, e.g., 1\,au. Thereafter, stellar fly-bys in the early stages might trigger a high-eccentricity migration. The young Hot Jupiter is created by tidal forces within a few Myrs. We expect those close-orbiting planets to have high inclinations. Highly aligned young planets, such as HIP 67522 b \citep{Rizzuto2020, Heitzmann2021}, are therefore probably not formed via high-eccentricity migration, even though the inclination evolution of planets in respect to the spin of the host-star can be complex \citep[e.g.,][]{Anderson2016, Vicketal2023}.  \\ 

\subsubsection{Examples for YHJ formation}
\label{Examples for YHJ formation}

\begin{figure}
\includegraphics[width=8cm]{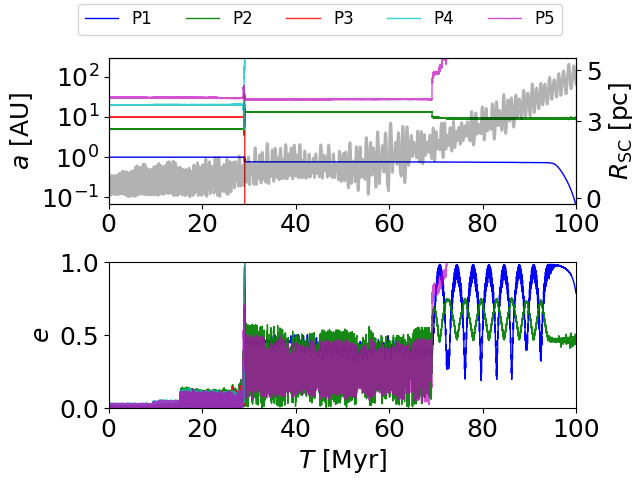}
\includegraphics[width=8cm]{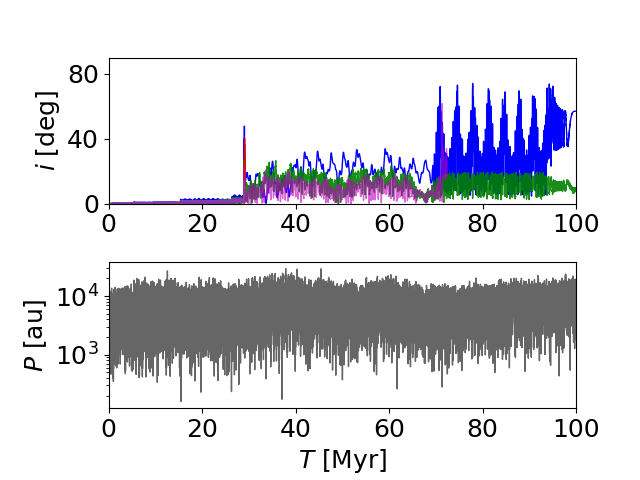}
\caption[Figure]{An example of the formation of a young Hot Jupiter in the planetary system \#119 (5PL1a model) in C64k star cluster. \\
\textit{Upper panel}: The colorful lines represent the planet's semi-major axes (upper subplot) and eccentricities (lower subplot). The  gray line in the upper subplot represents the distance to the center of the star cluster. \\
\textit{Lower panel}: The colorful lines represent the planet's inclinations to the initial plane (upper subplot). The black line depicts the distance to the closest perturber to the host star (lower subplot).}
\label{p_sys_119}
\end{figure}

\begin{figure}
\includegraphics[width=8cm]{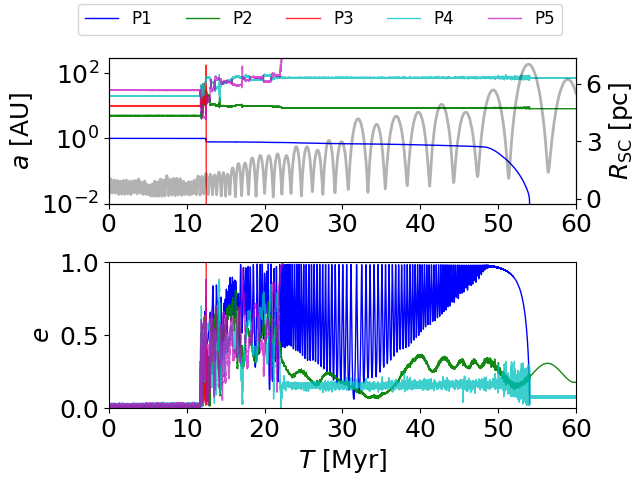}
\includegraphics[width=8cm]{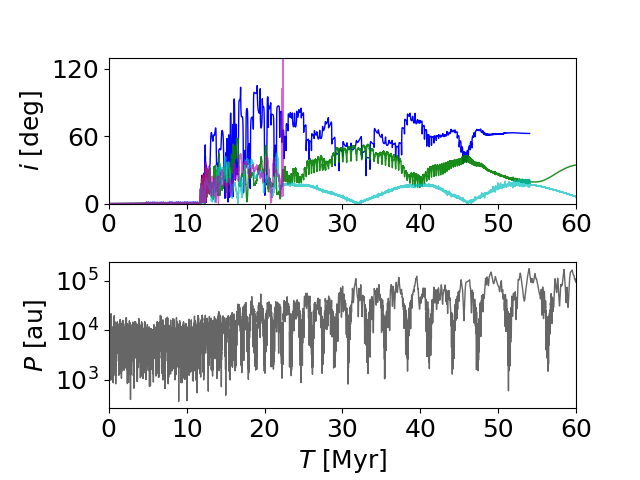}
\caption[Figure]{An example of young Hot Jupiter in the planetary system \#37 (5PL1a model) in the C32k star cluster. The young Hot Jupiter (blue line) is eventually tidally disrupted and excluded from the simulation.\\
\textit{Upper panel}: The colorful lines represent the planet's semi-major axes (upper subplot) and eccentricities (lower subplot). The gray line in the upper subplot depicts the distance to the center of the star cluster. \\
\textit{Lower panel}: The colorful lines represent the planet's inclinations to the initial plane (upper subplot). The black line depicts the distance to the closest perturber to the host star (lower subplot).}
\label{p_sys_37}
\end{figure}

Two examples of the formation of young Hot Jupiters are shown in Fig.~\ref{p_sys_119} and Fig.~\ref{p_sys_37}.  In Fig.~\ref{p_sys_119}, a strong encounter destabilizes the system and leads to the ejection of planets P3 and P4 after 30 Myrs.  However, the inner planet remains unaffected by the subsequent perturbations until another strong encounter occurs after 70 Myrs, causing it to reach higher eccentricities. The outer planet P5 is ejected within a few Myrs while P2 and P1 become trapped in a resonance. P1 undergoes a few resonance cycles with strong oscillations of eccentricity and inclination (but not a ZLK) until tidal forces eventually decouple it from the dynamic of the other planet and form a young Hot Jupiter. The simulation time of 100 Myrs ends before the young Hot Jupiter can reach circular orbit. \\
In Fig.\ref{p_sys_37}, we show a second example. A close encounter destabilizes the planetary system at roughly 12 Myrs and triggers a few hundred thousand years of planet-planet scattering. As a consequence, P3 is ejected and P1 is pushed to a semi-major axis of 0.78\,au. Over the following 10 million years, the outer planets change several times the position until also P5 is ejected. Mainly by angular momentum transfer with P2, the eccentricity of the innermost planet P1 reaches many times high values where tidal forces are effective but only for very short time intervals. At each eccentricity peak, tidal forces slightly shrink the semi-major axis of P1 and possibly prevent more extreme eccentricity values and tidal disruption. Finally, at approximately 48 Myrs, the planet reaches permanent eccentricity values $e > 0.95$. Tidal forces decouple the planet from the dynamics of the other planets and shrink the orbit to less than 10\,days within a few million years. As mentioned in Sec.~\ref{Tidal migration}, we have still not implemented planetary spin and Young Hot Jupiters, like this one, are eventually tidal disrupted and removed from the simulation. 
In both examples (although being in two different star cluster models), we observe a rough correlation between a host star's distance to the star cluster center and the average distance to the closest perturber. The number density of the stellar neighborhood surrounding a host star is significantly higher in the central regions of the star cluster than in the outskirts. When a host star resides in the central regions, the average distance to the closest perturber is relatively low, and the probability of a very close encounter (r $\leq$ 500 au) is comparatively high. Conversely, as the host star moves away from the star cluster center, both the average distance to the closest perturber and the likelihood of a very close encounter decrease. This indicates that the individual trajectory of a host star within a star cluster influences the stellar encounter rate and potentially affects the Hot Jupiter formation (see also Sec.~\ref{Constraints}).

\section{Discussion}
\label{Discussion}
\subsection{Comparison with previous studies}
\citet{Shara2016} used a bundle of N-body simulations to integrate the dynamics of 2 Jupiter-mass planet systems in a star cluster with 22\,000 members, dynamically evolved for 1 Gyr. The star cluster is less dense in the central regions compared to the C32k star cluster. It also has about 10\% primordial binaries compared to the total number of stars. Still, both star cluster environments are roughly comparable. Additionally, the two Jupiter-like planets have initial semi-major axes of 5.2\,au and 9.2\,au. They are located at a similar distance from their host star as our two inner planets in the 3PL5a and 4PL5a architecture.  However, with a HJC probability of 0.4\% per simulated planetary system, the results remain significantly below our found HJC probability of 1.5-4.5\% per simulated planetary system. Likely, the additional outer Jupiter-like planets at 20\,au and 30\,au enhance the probability of Hot Jupiter candidate formation even in relatively dense regions.  \\
As mentioned in Sec.~\ref{Direct scattering}, \citet{Hamersetal2017} found a Hot Jupiter candidate fraction of 0.02 in number densities of  $n_* \approx 10^4 \rm pc^{-3}$ for a Jupiter-like planet at a semi-major axis of 1\,au. The central densities of both our star cluster fall in this density range. Furthermore, our 5PL1a planetary model has the inner Jupiter-like planet at 1\,au. Nevertheless, at approximately 15\% of formed HJC, is far higher compared to the previous results of the authors. The results suggest that even if the inner planet is at 1\,au, mechanisms other than direct scattering can be more effective in creating HJ. If the planet's initial semi-major axis is initially at 5\,au, we witnessed no formation of an HJC via direct scattering. Thus, we can set the upper limit for the probability to have a HJC via direct scattering into $< 0.2$~\% per planetary system with planets initially beyond 5~au. This is consistent with the findings of \citet{Wang2020, Lietal2023}. \\
The methodology of a single stellar encounter on each planetary system of \citet{Wang2020, Wang2022} is different from our $N$-body approach using two codes without directly coupling them to each other. Therefore, a direct comparison between the two studies is difficult. However, our results confirm two of their findings: (i) The denser the star cluster and the further out a planet is orbiting, the higher the fraction of planetary systems with sufficient AMD, (ii) a multi-planet system with planets on further orbits contribute to the Hot Jupiter candidate formation preferentially in lower density regions.  \\ 
\subsection{Constraints on the formation of HJ}
\label{Constraints}

\begin{figure}
\includegraphics[width=8cm]{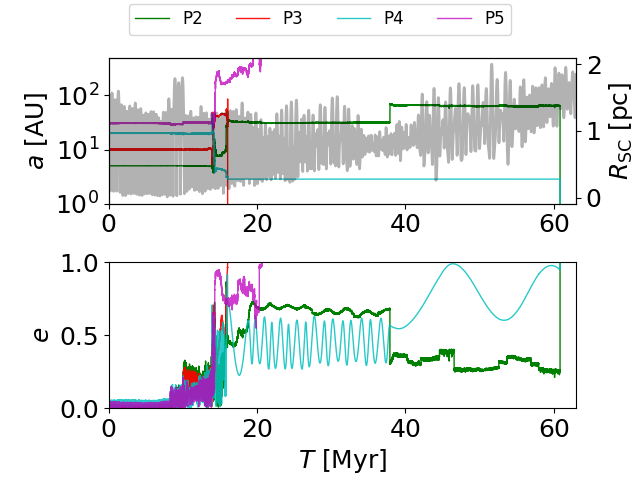}
\includegraphics[width=8cm]{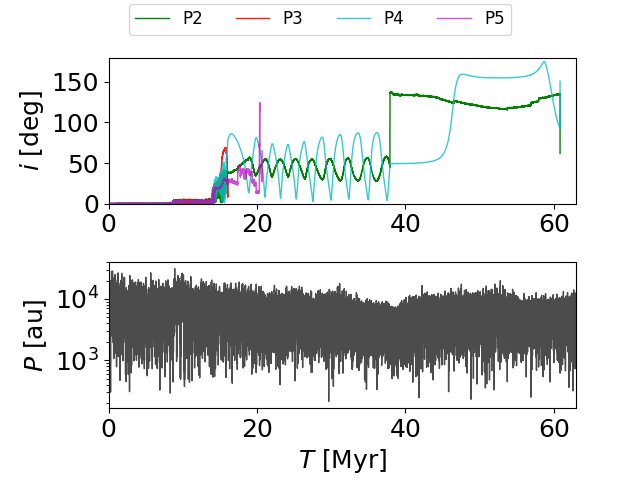}
\caption[Figure]{The dynamical evolution of the planetary system \#20 (4PL5a model) in C64k. This is an example of a HJC that does not become a Hot Jupiter because all planets are ejected by a stellar encounter (r $\leq$ 900\,au) after about 61 Myrs. \\
\textit{Upper panel}: The colorful lines represent the planet's semi-major axes (upper subplot) and eccentricities (lower subplot). The  gray line in the upper subplot depicts the distance to the center of the star cluster. \\
\textit{Lower panel}: The colorful lines represent the planet's inclinations to the initial plane (upper subplot). The black line depicts the distance to the closest perturber to the host star (lower subplot).}
\label{Failed_HJ}
\end{figure}

It is important to stress that in our tidal migration model, the majority of Hot Jupiter candidates will not become actual Hot Jupiters due to various factors, especially tidal disruption and further stellar encounters. Those constraints limit also the Hot Jupiter candidate formation itself.  \\
Tidal disruption occurs if the Hot Jupiter planet is pushed to a pericenter too close to the star ($p<r_{\rm{dis}}$, see Sec.~\ref{Tidal disruption}). Tidal dissipation itself does not significantly alter the pericenter \citep[e.g.,][]{Teyssandier2019}, but the planet may remain coupled to the dynamics of the rest of the system while under the influence of tidal forces. The angular momentum transfer of other planets can drive the Hot Jupiter candidate to higher and higher eccentricities. Ultimately, the planet may be pushed too close to the star and become tidally disrupted. \citet{Teyssandier2019, Vick2019} already labeled tidal disruption as a significant limitation to Hot Jupiter formation via secular chaos and proposed to consider in addition dynamical tides. If either (i) the effects of additional dynamical tides are taken into account or (ii) if the tidal disruption radius is smaller than the moderate value we have assumed, the number of tidal disrupted HJC could be limited. \\
The second main constraint is either (i) the destruction of the system, the ejection of planets, or (ii) general orbital parameters perturbation which alter the planetary system evolution. Notably, even larger distances of close stellar encounters, compared to the ones we showed in the examples, may perturb the planetary system and not result in the formation of an HJ. For stars with a few solar masses, fly-bys of several hundred\,au (e.g., a star with 4 Solar masses at r $\sim$ 900\,au in Fig.~\ref{Failed_HJ}) can be sufficient to change significantly the dynamics of the planetary system. Resonances such as the ZLK cycle are very sensitive to perturbation. Even a small change in orbital parameters of the outer planet can end the formation process of the Hot Jupiter. Also, the change of orbital parameters might lead to a pericenter of the inner planet too close to the star so that it is tidally disrupted. The result highlights the importance of considering multiple stellar encounters. Approaches with just one encounter and an unperturbed developing system thereafter may without further correction \citep[e.g., taking into account the expected survival rate,][]{Wang2022} be too simplistic to accurately estimate the Hot Jupiter fraction in star clusters. If tidal forces are stronger, the HJC decouples faster dynamics of the other planets and the circularisation time is smaller. Thus, stronger tidal forces could protect the HJC from the destructive influence of additional stellar encounters.   \\

\subsection{Estimation of actual Hot Jupiter fraction for 3/4PL5a}
  We provide a rough estimation of the expected Hot Jupiter fraction per planetary system ($f_{\rm HJ}$) for 4PL5a and 3PL5a with the innermost planet at 5 au, taking into account the aforementioned constraints. We assume the tidal forces taken in this paper (see Sec. \ref{Tidal migration}). With more effective tidal forces the transformation rate of HJC to actual Hot Jupiter could be significantly higher. To do this, we analyze each Hot Jupiter candidate individually at our final simulation time of 100 Myrs, considering their orbital parameters, AMD, and dynamical behavior over the previous Myrs. We also assume that no additional significant stellar encounters occur. We exclude systems where the Hot Jupiter candidate is unlikely to become an actual Hot Jupiter, such as those that have been tidally disrupted or ejected (see Fig.~\ref{Failed_HJ} for an example of such a system). \\ 
Our results suggest that extremely dense and slowly dissolving star clusters rather destroy the dynamic of systems with the innermost planet at 5\,au than create Hot Jupiters. In those star clusters – like the C64k – we expect a low fraction of systems with Hot Jupiters ($f_{\rm{HJ,4PL5a}} < 0.5 \%$) with a 4PL5a architecture. In initially less dense but faster-dissolving star clusters – like the C32k – the circumstances for Hot Jupiter formation seem to be much better. At least for the 4-planet system, the 4PL5a model, the probability of a system with a Hot Jupiter can be as high as  $f_{\rm{HJ,4PL5a}} \lessapprox 2 \%$. \\
The results for 3PL5a are ambiguous. We predict both star clusters to produce few Hot Jupiters ($f_{\rm{HJ, 3PL5a}} < 0.5 \%$). The dense C64k star cluster rather destroys the planetary system than forms a HJ. That does not seem to be the case for C32k. In the 4PL5a model in C32k, the HJ formation is efficient. Likely, the C32k does not provide enough perturbation. Outer orbiting planets are easier to be perturbed. Thus, a likely explanation for the low abundance of Hot Jupiters is the missing outer planet at 30\,au in the 3PL5a model compared to the 4PL5a model.  \\
Future work has to show if a wider 3-planet system can increase the fraction of systems with Hot Jupiters for the C32k star cluster or clusters with lower densities. It is possible that 3-planet systems with the initial semi-major axis of the outermost planet further out are more favorable in the Hot Jupiter production \citep[see][with the outer planet at 45-150\,au]{Wang2022}.

\section{Conclusion}
\label{Conclusion}
We presented numerical models of planetary systems subject to stellar encounters taken from a direct star cluster $N$-body model, and analyzed the probability of Hot Jupiter formation in them. The star cluster models had relatively high stellar density in their central region and 
Hot Jupiter formation occurred via high-eccentricity migration.
In every star cluster we inserted 200 identical planetary systems 
around Sun-like host stars. We use two different star cluster models with 32\,000 and 64\,000 members and central number densities of the order of $n_* \approx 10^4 \rm pc^{-3}$; there were three sets of planetary systems for each run. The planetary systems consist of 3, 4, or 5 Jupiter-mass planets amounting to a total of 200 planetary systems  times two star clusters with different densities times three different models of planetary systems. The 5-planet system model has the innermost planet at a semi-major axis of 1\,au whereas the 3- and 4-planet system has the innermost planet at 5\,au, as it is shown in Table \ref{PlanetsAU}. Our main results are listed in the following:  
\begin{itemize}
    \item  A combination of strong encounters, planet-planet scattering, and secular evolution within the planetary system frequently results in one of the planets becoming an HJC. The innermost planet has the highest probability to become a HJC, but there is a non-zero probability that the other planets also become HJ or HJC, according to the dynamical evolution history of the planetary system inside the star cluster.
    \item Systems with sufficient AMD for HJ formation via secular evolution ($C > C_r$, see Sec.~\ref{AMD}) are common in dense stellar environments. The fraction of planetary systems with sufficient AMD increases with higher densities and the presence of an additional outer planet ($a_4 = 30$ au). Contrarily, an additional inner planet ($a_1 = 1$ au) does not have a significant influence on the fraction of AMD-sufficient systems. In our limited sample size, no correlation between the fraction of AMD-sufficient systems and Hot Jupiter candidates (HJC) is witnessed. 
    \item The planetary systems of the 5PL1a model ($a_1 = 1$ au), are at 15\% the most likely to create Hot Jupiter candidates. If Jupiter-like planets are primordially formed or migrate at a semi-major axis of 1\,au, we expect Hot Jupiter to be more common in number density regions of $n_* \approx 10^4 \rm pc^{-3}$.
    \item Planetary systems with an additional outer planet at 30~au enhance HJ formation in comparably lower-density regions. However, in very dense star clusters, the additional outer planet is not an advantage in HJ production.
    \item Tidal disruption of the planet by the host star and additional close stellar encounters are the main constraints of the transformation of Hot Jupiter candidates to actual Hot Jupiters. Considering these limitations, we expect a maximum fraction of planetary systems ($a_1 = 5\,\mathrm{au}$) with a Hot Jupiter of 2\% with the assumed tidal forces. 
    \item Young Hot Jupiters ($t_{\rm age} < 100$ Myrs) can be formed via high-eccentricity migration if $a_1 = 1\,\mathrm{au}$. Those young Hot Jupiter are most likely misaligned. No young Hot Jupiter was witnessed in the 3PL5a and 4PL5a systems with $a_1= 5\,\mathrm{au}$. 
\end{itemize}

\section*{Acknowledgements}
We are grateful to the anonymous referee for providing comments and
suggestions that helped to improve this paper. Furthermore, we express our gratitude to Kai Wu for offering valuable feedback on the draft of this paper. 
LB, FFD and RS acknowledge support from the DFG priority program SPP 1992 ``Exploring the Diversity of Extrasolar Planets'' under projects Sp 345/20-1 and 22-1. We have used the ``Beijing'' version of \nb{}, which is supported by our team (see \cite{Kamlah2022a} and on \href{https://github.com/nbody6ppgpu/Nbody6PPGPU-beijing}{https://github.com/nbody6ppgpu/Nbody6PPGPU-beijing}).

%%%%%%%%%%%%%%%%%%%%%%%%%%%%%%%%%%%%%%%%%%%%%%%%%%
\section*{Data Availability}
The data underlying this article will be shared on reasonable
request to the corresponding author.



\bibliographystyle{mnras}
\bibliography{sgleo-hot-jupiters} % if your bibtex file is called example.bib

%\begin{thebibliography}{99}
%\end{thebibliography}

%%%%%%%%%%%%%%%%%%%%%%%%%%%%%%%%%%%%%%%%%%%%%%%%%%

%%%%%%%%%%%%%%%%% APPENDICES %%%%%%%%%%%%%%%%%%%%%

\appendix

 \section*{Appendix}
\renewcommand\thefigure{\thesection A.\arabic{figure}} 

\begin{figure}
\centering
\includegraphics[width=8cm]{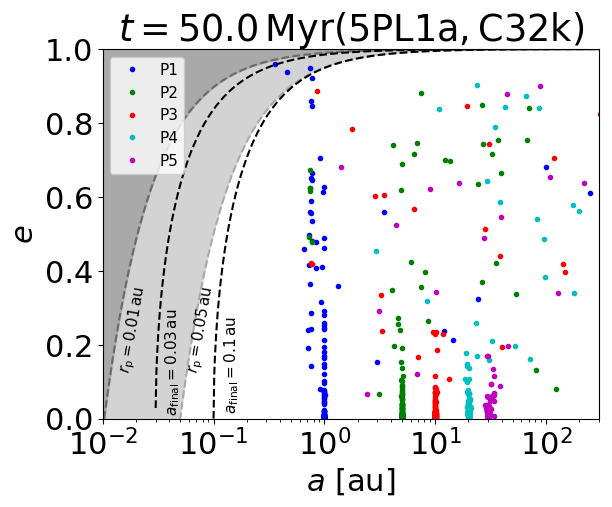}
\includegraphics[width=8cm]{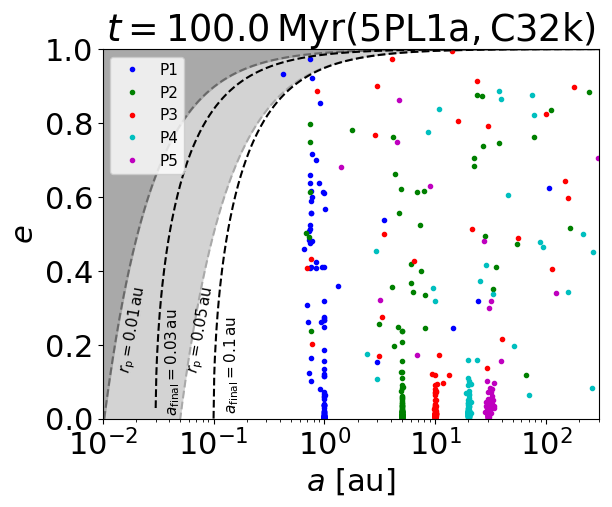}
\caption[Figure]{The semi-major axis-eccentricity parameter space of the 200 planetary systems of model 5PL1a in the C32k star cluster at 50 Myrs and at 100 Myrs. The color of the dots indicates a different planet. Each data point represents the planet's position in this parameter space at the time indicated. The dark gray area marks the region of the a-e space where planets can be tidally disrupted (i.e., $<0.01$~au), whereas the light gray area represents the region where tidal forces are effective ($<0.05$~au). The dashed lines represent the projected shrinking of the semi-major axis and eccentricity due to tidal forces with constant angular momentum (approximately following the constant time-lag theory, see Sec.~\ref{Tidal migration}) of a planet with a final semi-major axis of 0.03 and 0.1\,au.}
\label{a-e plot 5PL1a C32k}
\end{figure}

\begin{figure}
\centering
\includegraphics[width=8cm]{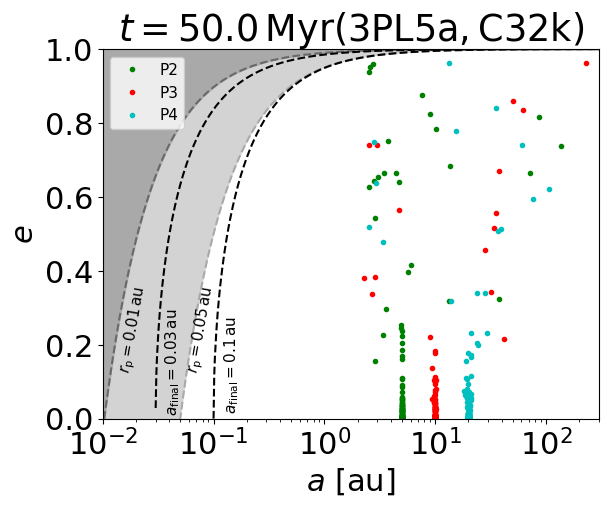}
\includegraphics[width=8cm]{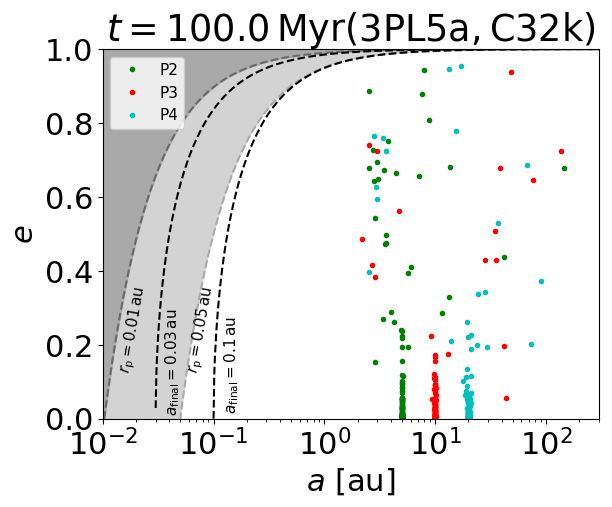}
\caption[Figure]{
Same as Figure \ref{a-e plot 5PL1a C32k}, but for model 3PL5a in the C32k star cluster at 50 Myrs and at 100 Myrs.
}
\label{a-e plot 3PL5a C32k}
\end{figure}

\begin{figure}
\centering
\includegraphics[width=8cm]{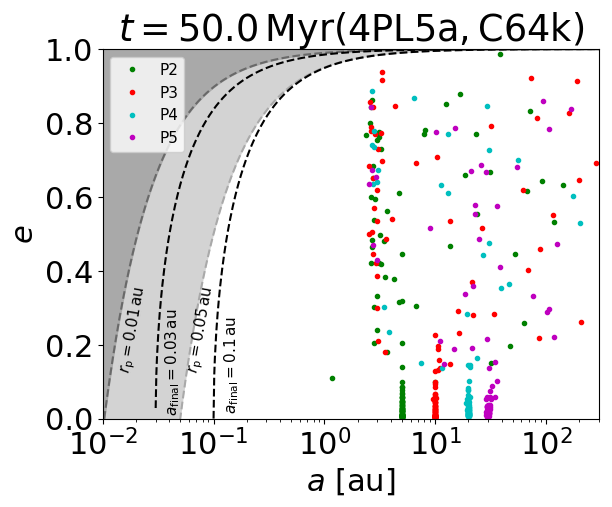}
\includegraphics[width=8cm]{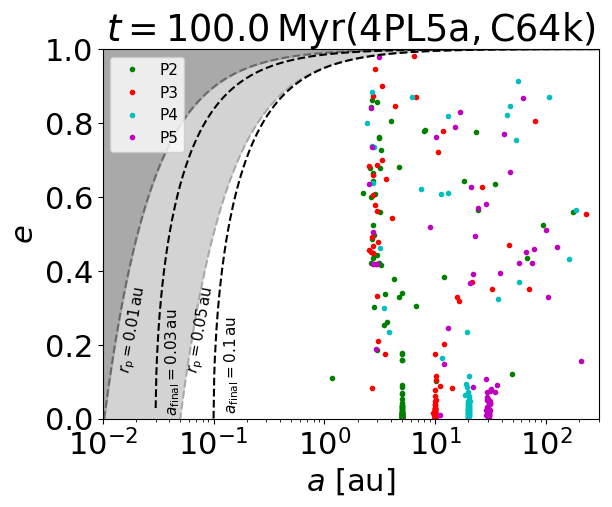}
\caption[Figure]{Same as Figure \ref{a-e plot 5PL1a C32k}, but for model 4PL5a in the C64k star cluster at 50 Myrs and at 100 Myrs.
}
\label{a-e plot else}
\end{figure}

\begin{figure}
\centering
\includegraphics[width=8cm]{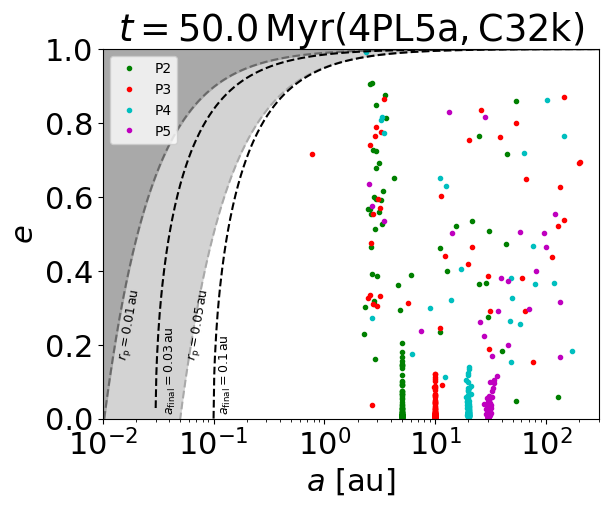}
\includegraphics[width=8cm]{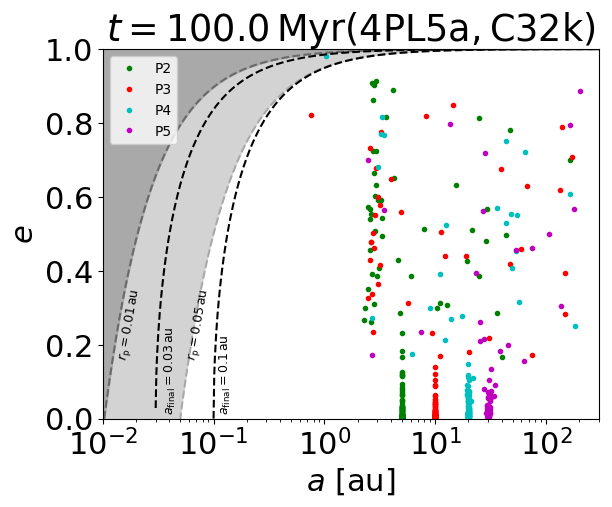}
\caption[Figure]{Same as Figure \ref{a-e plot 5PL1a C32k}, but for model 3PL5a in the C64k star cluster at 50 Myrs and at 100 Myrs.}
\label{a-e plot else}
\end{figure}

\begin{figure}
\centering
\includegraphics[width=8cm]{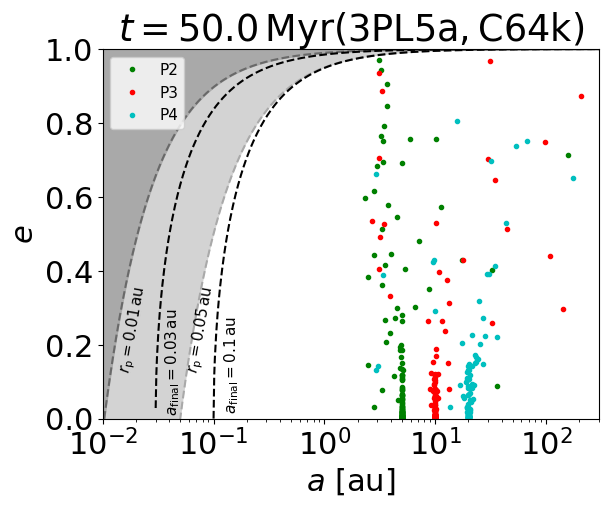}
\includegraphics[width=8cm]{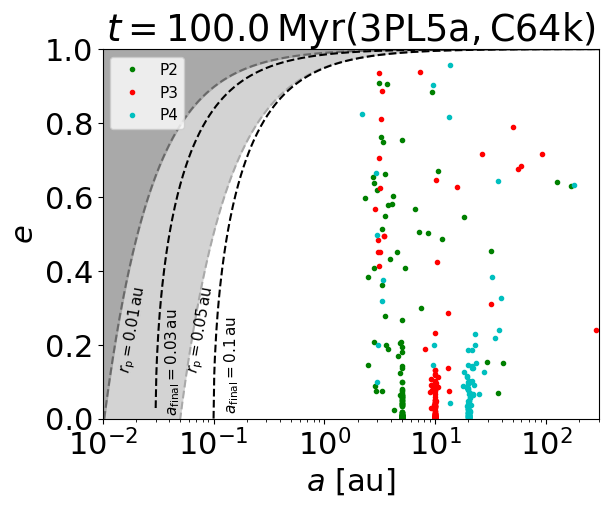}
\caption[Figure]{Same as Figure \ref{a-e plot 5PL1a C32k}, but for model 3PL5a in the C64k star cluster at 50 Myrs and at 100 Myrs.}
\label{a-e plot else}
\end{figure}

\begin{figure}
\label{}
\includegraphics[width=8cm]{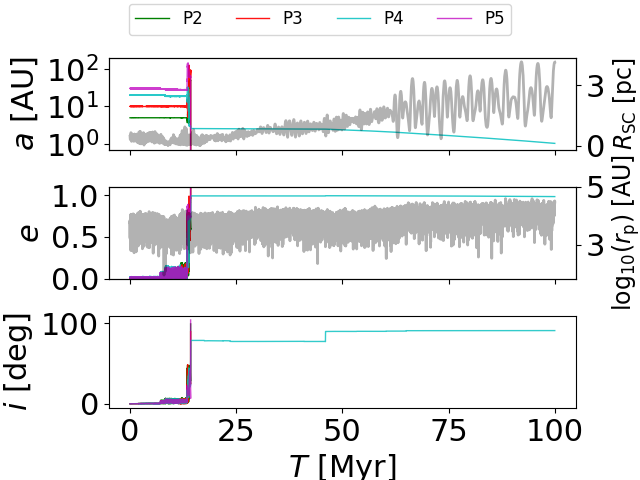}
\caption[Figure]{Evolution of \#65 4PL5a model in the C32k star cluster. The colorful lines represent the planet's semi-major axes (upper subplot), eccentricities (middle subplot), and inclinations (lower subplot) whereas the gray lines represent the distance to the center of the star cluster (upper subplot) and the distance to the closest perturber (middle subplot). After 14 Myrs a stellar encounter (r $\leq$  71\,au) followed by very short but strong planet-planet scattering P3 is perturbed to eccentricity values e $\sim$ 0.99  and labeled as a HJC. At 46 Myrs, another encounter accelerates the circularization and the semi-major axis shrinks notably over the next tens of millions of years from approximately 2.5 to 1.0\,au.  We expect P3 to become a Hot Jupiter within the next 50 Myrs.}
\end{figure}

\begin{figure}
\label{}
\includegraphics[width=8cm]{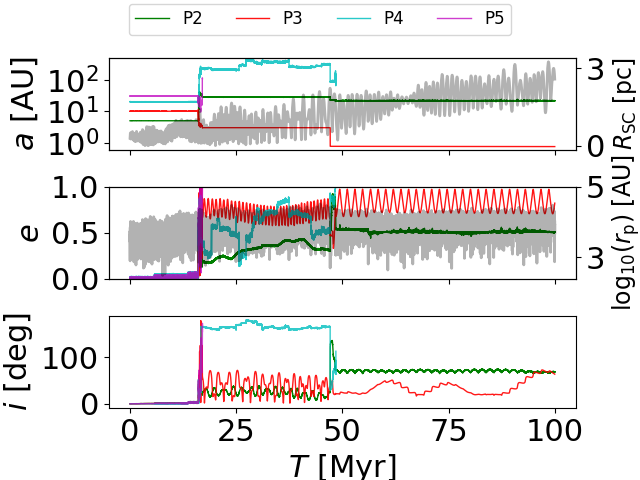}
\caption[Figure]{Evolution of \#170 4PL5a model in the C32k star cluster. 2 strong stellar encounters at 16 Myrs and 47 Myrs bring planet P3 to a semi-major axis of 0.78\,au. Then, it reaches high-eccentricity peaks roughly every 2 million years and is classified as a HJC.}
\end{figure}

\begin{figure}
\label{}
\includegraphics[width=8cm]{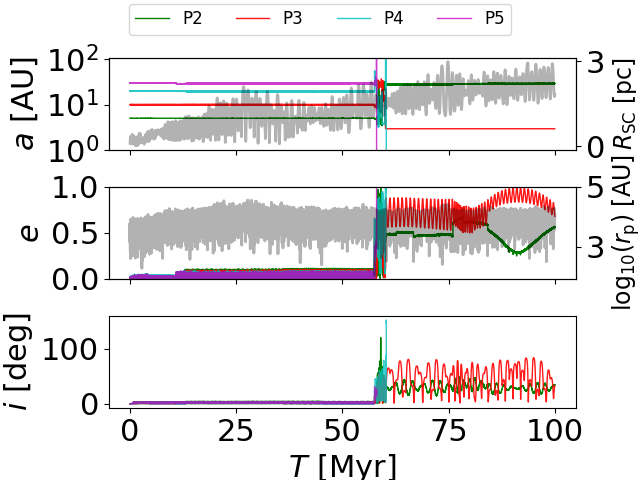}
\caption[Figure]{Evolution of \#113 4PL5a model in the C32k star cluster. A strong stellar encounter after 58 Myrs followed by planet-planet scattering ejects P4 and P5. The remaining P3 and P2 are influenced by additional stellar encounters and influence each other. At roughly 90 Myrs P3 reaches high eccentricity values (e $\sim$ 0.99) and is classified as a HJC.}
\end{figure}

\begin{figure}
\label{}
\includegraphics[width=8cm]{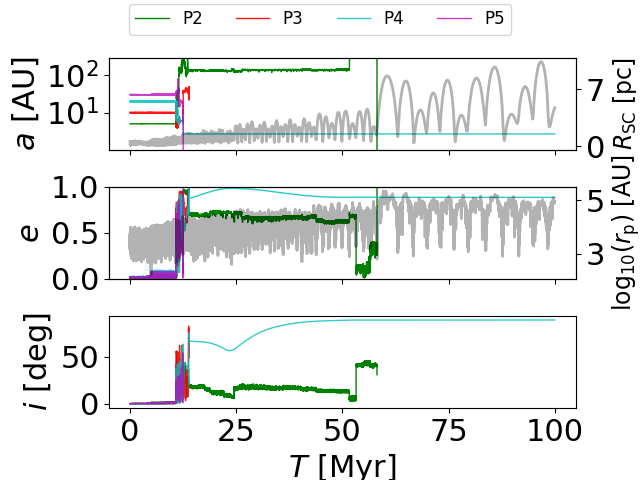}
\caption[Figure]{Evolution of \#115 4PL5a model in the C64k star cluster. At 11 Myrs a stellar encounter and a short period of strong planet-planet scattering P4 is pushed to a semi-major axis of roughly 2.6\,au and reaches high eccentricities (e $\sim$ 0.98) after 20 Myrs and is classified as an HJC. However, a strong encounter influences the now outer planet P1 and stops the formation process}
\end{figure}

\begin{figure}
\label{}
\includegraphics[width=8cm]{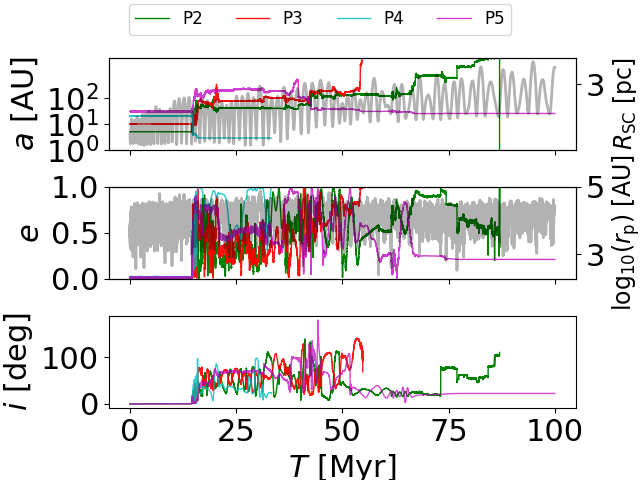}
\caption[Figure]{Evolution of \#9 4PL5a model in the C32k star cluster. At about 17 Myrs a close encounter destabilizes the system. After the encounter and planet-planet scattering, the closest planet to the star is P3. In the following roughly 20 million years the system undergoes a chaotic evolution and P3 reaches high eccentricities of e $=$ 0.99 where tidal forces are effective. However, at 32 Myrs the planet is driven too close to the star (periastron $\sim$ 0.09\,au) by the dynamics of other planets and is tidally disrupted and removed from the system.}
\end{figure}

%%%%%%%%%%%%%%%%%%%%%%%%%%%%%%%%%%%%%%%%%%%%%%%%%%

% Don't change these lines
\bsp	% typesetting comment
\label{lastpage}
\end{document}